\documentclass[12pt]{article}

\usepackage{xcolor}
\usepackage{multirow}
\usepackage{comment}
\usepackage{epsfig}
\usepackage{epstopdf}
\usepackage{amssymb,amsmath}
\usepackage{cite}

\usepackage{setspace}
\newcommand{\bea}{\begin{eqnarray}}
\newcommand{\eea}{\end{eqnarray}}
\newcommand{\gray}[1]{\textcolor{gray}{#1}}

\numberwithin{equation}{section}

\begin{document}
\begin{titlepage}
%\begin{flushright}
%
%\end{flushright}
%
\vspace*{10mm}
\begin{center}
\baselineskip 25pt 
{\Large\bf
%%%%%%%%%%%%%%%%%%%%%%%%%%%%%%%%%%%%%%%%%%%%%%%%%%%
Axions, WIMPs, proton decay and observable $r$ in $SO(10)$ 
%%%%%%%%%%%%%%%%%%%%%%%%%%%%%%%%%%%%%%%%%%%%%%%%%%%
}
\end{center}
\vspace{5mm}
\begin{center}
{\large
Nobuchika Okada\footnote{okadan@ua.edu}, 
Digesh Raut\footnote{draut2@washcoll.edu}, 
and Qaisar Shafi\footnote{qshafi@udel.edu}
}
\end{center}
\vspace{2mm}

\begin{center}
{\it
$^{1}$ Department of Physics and Astronomy, \\ 
University of Alabama, Tuscaloosa, Alabama 35487, USA \\
$^{2}${Physics Department,  Washington College,  Chestertown,  MD 21620, USA}\\
$^{3}$ Bartol Research Institute, Department of Physics and Astronomy, \\
 University of Delaware, Newark DE 19716, USA
}
\end{center}
\vspace{0.5cm}
%%%%%%%%%%%%%%%%%%%%%%
\begin{abstract}
%%%%%%%%%%%%%%%%%%%%%%
We explore some experimentally testable predictions of an
$SO(10)$ axion model which includes two 10-plets of fermions in order to resolve the axion domain wall problem. The axion symmetry can be safely broken after inflation, so that the isocurvature perturbations associated with the axion field are negligibly small. An unbroken gauge $Z_2$ symmetry in $SO(10)$ ensures the presence of a stable WIMP-like dark matter, a linear combination of the electroweak doublets in the fermion 10-plets and an $SO(10)$ singlet fermion with mass $\sim 62.5 \; {\rm GeV}\; (1 \; {\rm TeV}) $ when it is mostly the singlet (doublet) fermion, that co-exists with axion dark matter. We also discuss gauge coupling unification, proton decay, inflation with non-minimal coupling to gravity and leptogenesis. 
With the identification of the SM singlet Higgs field in the $126$ representation of $SO(10)$ as inflaton, the magnetic monopoles are inflated away, and we find $0.963 \lesssim n_s \lesssim 0.965$ and $0.003 \lesssim r \lesssim 0.036$, where $n_s$ and $r$ denote the scalar spectral index and tensor-to-scalar ratio, respectively. 
These predictions can be tested in future experiments such as CMB-S4. 

%and the tensor to scalar ratio r, a canonical measure of gravity waves from inflation, is estimated to be $r \simeq 0.003$, which can be tested in future experiments such as CMB-S4. 
%%%%%%%%%%%%%%%%%%%%%%%
\end{abstract}
\end{titlepage}

%%%%%%%%%%%%%%%%%%%%%%%%%%%%%%%%%
\section{Introduction}
\label{sec:Intro}
%%%%%%%%%%%%%%%%%%%%%%%%%%%%%%%%%
%Aside from the fact that the Standard Model (SM) of particle physics does not account for various well-established experimental observations in both cosmology and particle physics, there are also sound theoretical motivations/reasoning to extend the SM with new physics beyond the SM, as we discuss below. 
It is well-known that the Standard Model (SM) of particle physics needs to be extended to account for various well-established experimental observations in both cosmology and particle physics that the SM fails to explain.  
To do so, one may extend the symmetry of the SM with new particles. 
To solve the so-called strong CP problem \cite{Peccei:2006as} originating from the observed tiny electric dipole moment of neutron \cite{Afach:2015sja}, Peccie and Quinn (PQ) introduced a global $U(1)_{PQ}$ symmetry. 
Interestingly, the pseudo-Goldston boson (axion) associated with the spontaneous breaking of this $U(1)_{PQ}$ symmetry is a potential candidate for dark matter (DM) \cite{Weinberg:1977ma}.  
About 25\% of the total energy density in the universe is attributed to the DM \cite{Aghanim:2018eyx}.

To be phenomenologically viable, this axion scenario must overcome two major cosmological hurdles, namely, the axion-domain-wall problem and the isocurvature problem (for a review, see, for example, Ref.~\cite{Kawasaki:2013ae}). 
If the domain-wall number associated with the PQ symmetry breaking, $N_{DW}>1$, which is typically the case, the topological defects (strings and domain-walls) produced during this PQ phase transition dominate the energy density of the present universe, which is not consistent with the observation. 
An inflation scenario, which takes place after the PQ symmetry breaking, is often invoked to solve the domain-wall problem. 
However, this leads to large isocurvature perturbations, which are severely constrained by the Planck observations.\footnote{To suppress the generation of axion isocurvature, the value of Hubble parameter during inflation should be $H_{inf} \lesssim 10^{7}$ GeV, which is much smaller than $H_{inf} \simeq 10^{13-14}$ GeV that a simple Higgs inflation scenario can realize (see, for example, Ref.~\cite{Shafi:2006cs, Okada:2010jf, Okada:2014lxa}). An inflaton potential that exhibits an approximate inflection-point during inflation \cite{Okada:2016ssd} realizes such low Hubble values and has been implemented in (GUT) models with axions in Refs.~\cite{axionIPI, IPISO10}.} 
In this work, we consider the setup with the domain-wall number $N_{DW}=1$, for which the domain-wall rapidly decays and contributes negligibly to the total energy density of the present universe \cite{Kawasaki:2013ae}. 
With the domain-wall problem resolved, there is no isocurvature problem if the PQ symmetry breaking takes place after the end of inflation.

A domain-wall free axion model in the context of grand unified theories (GUTs) was discussed in Ref.~\cite{Kibble:1982ae}. 
Such a model based on $SO(10)$ GUT which is broken to the SM gauge group, $SU(3)_c \times SU(2)_L \times U(1)_Y$, 
via the Pati-Salam (PS) gauge group \cite{Pati:1974yy}, $SU(4)_c \times SU(2)_L\times SU(2)_R$, was proposed in Ref.~\cite{Holman:1982tb} (see also Ref.~\cite{Ernst:2018bib}). 
In addition to three generations of fermions in the ${\bf 16}$ representation of $SO(10)$ (each ${\bf 16}$-plet includes all the SM fermions in a corresponding generation and an SM singlet right-handed neutrino), two fermions in the ${\bf 10}$ representation of $SO(10)$ are introduced. 
The ${\bf 16}$-plet and the ${\bf 10}$-plet fermions are assigned PQ charges $+1$ and $-2$, respectively. 
With this charge assignment, after the $SO(10)$ symmetry breaking, the PQ symmetry is broken down to ${\bf Z_4}$, which is identified with the center of $SO(10)$ (more precisely, $Spin (10)$). 
Equivalently, the domain-wall number $N_{DW}=1$ and there is no domain-wall problem. 
However, both the $SO(10)$ and PS symmetry breakings copiously produce point topological defects called monopoles with masses of the order of the symmetry breaking scales \cite{monopole1}. 
The monopoles created by the $SO(10)$ (PS) symmetry breaking at $M_{GUT}\simeq 10^{15}$ ($M_{PS} \simeq 10^{11}$) GeV will dominate the energy density of the universe \cite{monopole2,monopole3}, which is not consistent with the observation.  
This so-called monopole problem was not considered in Ref.~\cite{Holman:1982tb}.

One of the original motivations of the cosmological inflation scenario proposed by Guth \cite{Guth} was to solve the monopole problem by diluting the monopole density. 
If the value of the Hubble parameter during the inflation $H_{inf}\ll M_{GUT, PS}$, both the GUT and the PS monopole problem can be solved. 
Simple inflation scenarios predicting $H_{inf} \simeq 10^{13-14}$ GeV \cite{Okada:2014lxa} cannot solve the monopole problem arising from the PS symmetry breaking.\footnote{The lighter monopoles produced during the symmetry breaking could survive inflation for $M_{PS} \lesssim 10^{13-14}$ GeV and may exist at an observable level \cite{Senoguz:2015lba, Chakrabortty:2020otp}.} 
However, if the inflaton is a Higgs field that breaks the PS symmetry, the monopole problem can be solved even if $H_{inf} > M_{PS}$. 
This is because the PS symmetry is already broken during the inflation.

In this paper, we extend the $SO(10) \times U(1)_{PQ}$ model proposed in Ref.~\cite{Kibble:1982ae} with one $SO(10)$ singlet fermion with PQ charge $+4$. 
Since the new particle is an $SO(10)$ singlet, the axion model structure remains the same as the original model. 
In the presence of the singlet fermion, a charge-neutral linear combination of the SM doublet field in the ${\bf 10}$-plet and the singlet fermion, which is a Majorana fermion, serves as another DM candidate. 
This is called the singlet-doublet fermion DM. 
Note that in the absence of the singlet fermion, because of the PQ symmetry, the 10-plet fermion is a stable Dirac fermion DM candidate with weak interactions.  
However, such a weakly interacting Dirac fermion DM must be very heavy, $m_{DM} \gtrsim 10^9$ GeV \cite{Dunsky:2020yhv, Okada:2021uqk} to avoid very severe constraints from the direct DM detection experiments \cite{XENON:2018voc, Akerib:2018lyp}. 
This DM mass scale is far above the so-called unitarity bound \cite{Griest:1989wd}.

After imposing various phenomenological constraints, including the requirement of a successful unification of the gauge couplings, we identify the allowed region of the model parameter space. 
We find that in most of the allowed parameter space, the axion decay constant $f_a < 7.11 \times 10^{11}$ GeV, for which the axion cannot account for 100\% of the observed DM abundance in the universe \cite{Kawasaki:2013ae}. 
Thus, most of the observed DM is the fermion DM with  mass $\sim 62.5 \; {\rm GeV} \;(1 \; {\rm TeV}) $ when it is mostly the singlet (doublet) fermion. 
We find that $f_{a}$ can be as low as $10^8$ GeV which is relevant to the solar axion interpretation of the excess in the electron recoil events recently reported by XENON1T experiment \cite{Xenon1t1}. 
Although such a small $f_a$ value is in strong tension with the stellar cooling constraints \cite{Xenon1t1}, subsequent analysis has shown the tension can be significantly relaxed. See, for example, Ref.~\cite{Xenon1t3}.

We also consider inflation scenario in our model. 
We identify a PS symmetry breaking Higgs field to be the inflaton with a non-minimal gravitational coupling with a coupling constant $\xi$.  
We find that a successful unification of the gauge couplings requires the inflaton quartic coupling values to be $\sim 0.1$ or greater, which corresponds to $\xi \gtrsim 10^4$. 
The corresponding inflationary predictions are consistent with the current observations and can be tested in future CMB experiments. 
The inflaton vacuum expectation value generates the Majorana masses for the right-handed neutrinos which play a crucial role to explain the observed light neutrino masses via the type-I seesaw mechanism \cite{Seesaw}. 
Focusing on this coupling between the inflaton and right-handed neutrinos, we discuss the reheating from the inflaton decay to right-handed neutrino pairs. 
We identify an upper bound on right-handed neutrino Yukawa couplings to reheat the universe to a sufficiently high temperature that allows for a successful leptogenesis \cite{leptogenesis1} but which is lower than $M_{PS}$ to prevent the restoration of the PS symmetry.

The rest of this paper is organized as follows: we introduce our model in Sec~\ref{sec:model}. 
In Sec.~\ref{sec:GCUandPD} we identify the region in the model parameter space to realize a successful unification of gauge couplings which is consistent with various phenomenological bounds, including bounds from proton lifetime measurements. 
We show that this will constrain the allowed range for the PQ symmetry breaking scale. 
In Sec.~\ref{sec:inf} we discuss the non-minimal inflation scenario with the identification of the SM singlet scalar field in the PS symmetry breaking Higgs multiplet to be the inflaton. 
We also identify the model parameter region to realize a successful reheating scenario and leptogenesis. 
In Sec.~\ref{sec:DM} we discuss the axion and the singlet-doublet DM scenario. 
For either one of these or their combination to reproduce the observed relic abundance of the DM, we identify the model parameter region which is consistent with various experimental bounds. 
Our conclusions are summarized in Sec.~\ref{sec:conc}.

%%%%%%%%%%%%%%%%%%%%%%%%%%%%%%%%%%%%%%
\section{$SO(10) \times U(1)_{PQ}$ Model}
\label{sec:model}
%%%%%%%%%%%%%%%%%%%%%%%%%%%%%%%%%%%%%%

The particle content of the $SO(10) \times U(1)_{PQ}$  model is listed in Table~\ref{tab:PC}.  
The fermions are in the ${\bf 16}$ ($+1$), ${\bf 10}$ ($-2$), and ${\bf 1}$ ($+4$) representations of  $SO(10) \times U(1)_{PQ}$. 
Each of the three ${\bf 16}$-plet fermions, ${16}_{SM}^{(i = 1,2,3)}$, includes one generation of SM fermions plus a new SM singlet right-handed neutrino. 
The rest are the Higgs (scalar) fields which break $SO(10) \times U(1)_{PQ}$ to the gauge symmetry $SU(3)_c \times U(1)_{EM}$ at low energies. 

%%%%%%%%%%%%%%%%%%%%%%%%%%%%%%%%%%%%%%
\begin{table}[t]
\begin{center}
\begin{tabular}{ll|c|c|c}
\textbf{}                      & & $SO(10)$ & $U(1)_{PQ}$   \\ \hline
\multicolumn{1}{l|}{\multirow{3}{*}{\textbf{Fermions}}} 
&$ { 16}_{SM}^{(i)} $  &  {\bf 16}   & + 1  \\
\multicolumn{1}{l|}{}
&$ {10}_{E}^{(j)} $    & {\bf 10}   & -- 2  \\ 
\multicolumn{1}{l|}{}                                   
&$ {1}_{E} $    &  {\bf 1} & + 4   \\ \hline
\multicolumn{1}{l|}{\multirow{6}{*}{\textbf{Scalars}}}  
&$ {10}_{H} $    &  {\bf 10}   & -- 2  \\
\multicolumn{1}{l|}{}                                   
&$ {45}_{H} $    & {\bf 45}   &   + 4  \\
\multicolumn{1}{l|}{}                                   
&$ {126}_{H} $    &  {\bf 126}  & + 2  \\
\multicolumn{1}{l|}{}                                   
&$ {210}_{H} $    &  {\bf 210}   & \;\;\;0  \\ 
\multicolumn{1}{l|}{}                                   
&$ {\Phi} $    &  {\bf 1}   & -- 8 \\
\hline 
\end{tabular}
\end{center}
\caption{
Particle contents of $SO(10) \times U(1)_{PQ}$ model.
%: the fermion sector includes three ($i = 1,2,3$) {\bf 16}-plet fermions, each of which includes all the SM fermions in a single generation plus a SM singlet Majorana fermion, two ($j = 1,2$) {\bf 10}-plet fermions, and a $SO(10)$ singlet fermion.   The rest are the scalar fields which break $SO(10) \times U(1)_{PQ}$ to the SM. 
}
\label{tab:PC}
\end{table}
%%%%%%%%%%%%%%%%%%%%%%%%%%%%%%

To discuss the symmetry breaking, we list the decomposition of the various Higgs representations under the PS gauge group, 
    $SU(4)_c \times SU(2)_L\times SU(2)_R$,      
\bea
{\bf 210} &=& ({\bf 1},{\bf 1},{\bf 1}) \oplus ({\bf 15},{\bf 1},{\bf 1}) \oplus ({\bf 6}, {\bf 2},{\bf 2}) \oplus ({\bf 15},{\bf 3},{\bf 1}) \oplus ({\bf 15},{\bf 1},{\bf 3}) \oplus ({\bf 10},{\bf 2},{\bf 2}) \oplus (\overline{\bf 10},{\bf 2},{\bf 2}) , \nonumber \\
{\overline {\bf 126}} &=& (\overline{\bf 10},{\bf 1},{\bf 3}) \oplus ({\bf 15}, {\bf 2}, {\bf 2}) \oplus ({ {\bf 10}}, {\bf 3}, {\bf 1}) \oplus ({\bf 6},{\bf 1},{\bf 1}), \nonumber \\
{\bf 45} &=& ({\bf 1},{\bf 1},{\bf 3}) \oplus ({\bf 15},{\bf1},{\bf1}) \oplus ({\bf 1},{\bf 3},{\bf 1}) \oplus ({\bf 6},{\bf 2},{\bf 2}), \nonumber \\
{\bf 10}&=& ({\bf 1},{\bf 2},{\bf 2}) \oplus ({\bf 6},{\bf 1},{\bf 1}). 
\label{eq:PSdecomp}
\eea 
Below we outline the symmetry breaking pattern with the Higgs fields and their VEVs at each step of the breaking, 
\bea
SO(10) \times U(1)_{PQ}
& \xrightarrow{\langle{210}_{H}\rangle}&
SU(4)_c \times SU(2)_L \times SU(2)_R \times U(1)_{PQ} \nonumber \\
& \xrightarrow[]{\langle{\overline {126}}_{H}\rangle}&
SU(3)_c \times SU(2)_L \times U(1)_{Y} \times U(1)^\prime_{PQ}
\nonumber \\
& \xrightarrow[]{\langle{45}_{H}\rangle, \; \langle{ \Phi}\rangle}& 
SU(3)_c \times SU(2)_L \times U(1)_{Y}  \nonumber \\
& \xrightarrow {\langle{10}_{H}\rangle}& 
SU(3)_c \times U(1)_{EM}. 
\label{eq:SB}
\eea
We assume a suitable Higgs potential in this paper in order to implement the desired symmetry breaking pattern. 
The $SO(10)$ gauge symmetry is spontaneously broken to the intermediate PS gauge group when the PS singlet component of $210_H$, $({\bf 1},{\bf 1},{\bf 1})$, develops a GUT scale VEV, $ \langle {210}_{H} \rangle = M_{\rm GUT}$. 
After the $(\overline{\bf 10},{\bf 1},{\bf 3})$ component of ${\overline {126}}_{H}$ develops a VEV $\langle{\overline {126}}_{H}\rangle = M_{PS}$, the PS gauge group is broken down to the SM gauge group. 
A global $U(1)^\prime_{PQ}$ survives this breaking, which is a linear combination of the original PQ symmetry and the $B-L$ symmetry \cite{Kibble:1982ae, Holman:1982tb, Lazarides:2020frf}.  
This global symmetry is broken after the $({\bf 15},{\bf1},{\bf1})$  and $({\bf 1},{\bf1},{\bf3})$ components of ${45}_{H}$ and the SM singlet scalar ${\Phi}$ develop their VEVs, namely, 
 $\langle{45}_{H}\rangle = v_{45}$ and $\langle{\Phi}\rangle = v_{\Phi}$. 
In our analysis we simply set $v_{45} = v_{\Phi}$ such that $f_a \simeq v_{45}/N_{DW} = v_{45}$ (up to ${\cal O}(1)$ Clebsch-Gordan coefficient and $N_{DW} = 1$). 
As we have discussed earlier, it is also crucial that the $U(1)^\prime_{PQ}$ remains unbroken during the inflation to avoid the generation of isocurvature perturbations from the fluctuation of PQ/axion field. 
To realize this situation, we assume a mixed quartic coupling of $45_H$ with the inflaton field which induces a large positive mass squared for $45_H$ during inflation to keep $U(1)^\prime_{PQ}$ unbroken.
The SM gauge symmetry obtained after the breaking of $U(1)^\prime_{PQ}$ breaks to $SU(3)_c \times U(1)_{EM}$ at the electroweak scale when the $SU(2)_L$ doublet Higgses, two in ${10}_{H}\supset ({\bf 1},{\bf 2},{\bf 2})$ and the other two in ${\overline {126}}_{H} \supset ({\bf 15}, {\bf 2}, {\bf 2})$, all develop the electroweak scale VEV.\footnote{
The electroweak scale VEV for the $({\bf 15}, {\bf 2}, {\bf 2})$ can be realized by an induced VEV mechanism from a mixed scalar coupling ${126} \; {\overline {126}}\;  {126} \; {10}_H$ \cite{Lazarides:1980nt}.}

Except for those components of scalars listed in Eq.~(\ref{eq:PSdecomp}) which are involved in the symmetry breaking, all the other scalar components are assumed to have GUT scale masses. 
We will discuss the masses of $(\overline{\bf 10}, {\bf 1}, {\bf 3}) \subset {\overline {126}}_{H}$ and $45_H$ in Sec.~\ref{sec:GCUandPD}. 
For the four $SU(2)_L$ doublet Higgs fields in $({\bf 1},{\bf 2},{\bf 2}) \subset{10}_{H}$ and  $({\bf 15}, {\bf 2}, {\bf 2}) \subset {\overline {126}}_{H}$, we assume that only one linear combination lies at the electroweak scale, which we identify to be the SM Higgs doublet, while the rest of the linear combinations are assumed to be at the PS scale.

The Yukawa interactions for the SM fermions and the three right-handed neutrinos are as follows:  
\bea
 {\cal L} \supset  {16^i}_{SM} \left(Y_{10}^{ij}  {10}_{H} + \frac{1}{4}Y_{\overline{126}}^{ij} {\overline {126}}_{H}  \right)  {16}_{SM}^j, 
\label{eq:SMY}
\eea
where $i,j$ are the generation indices. 
Note that the Yukawa term ${16}_{SM} {10}_{H}^* {16}_{SM}$ is forbidden by the PQ symmetry.
This so-called minimal $SO(10)$ model can reproduce realistic mass matrices for the SM fermions.  
See, for example, the papers listed in Ref.~\cite{S010nonSUSY}. 
A detailed analysis of the SM fermion masses is beyond the scope of this work. 
The Yukawa couplings of the new fermions are given by 
\bea
 {\cal L} \supset  \sum_{i \neq j} \frac{1}{2}Y_{45}^{(ij)} {45}_H  {10}_{E}^{(i)} {10}_{E}^{(j)}  + \sum_{j} { Y_H}^{(j)} {10}_{H} { 10}_{E}^{(j)}  {1}_{E} +  \frac{1}{2} Y_{\Phi} {\Phi}  { 1}_{E} { 1}_{E}, 
\label{eq:ExoticY}
\eea 
where $Y_{45}^{(ij)}$ is anti-symmetric.  
We will discuss their mass spectrum in the next section.

%%%%%%%%%%%%%%%%%%%%%%%%%%%%%%%%%%%%%%
\section{Gauge Coupling Unification and Proton Decay}
\label{sec:GCUandPD}
%%%%%%%%%%%%%%%%%%%%%%%%%%%%%%%%%%%%%  
The unification of the SM gauge couplings is realized in two steps. 
Starting from low energies, the SM gauge couplings link with the PS gauge couplings at the intermediate scale $M_{PS}$ followed by the unification of the PS couplings at $M_{GUT}$. 
After determining the mass spectrum of the new particles that contribute to the renormalization group (RG) running of the gauge couplings, 
we will identify the parameter space to realize a successful unification of the gauge couplings and also be consistent with various phenomenological constraints.

%%%%%%%%%%%%%%%%%%%%%%%%%%%%%%%%%%%%%%
\subsection{New Particle Mass Spectrum}
\label{sec:MassSpec}
%%%%%%%%%%%%%%%%%%%%%%%%%%%%%%%%%%%%% 
In this subsection we consider the new fermions and scalars with masses smaller than the PS symmetry breaking scale.
The relevant components of the Higgs fields in ${\bf 126}$ and ${\bf 45}$ representations of $SO(10)$ in Eq.~(\ref{eq:PSdecomp}) decompose under the SM gauge group, $SU(3)_c \times SU(2)_L \times U(1)_Y$ as  
\bea
({\bf 10},{\bf 1},{\bf 3}) &=& ({\bf1},{\bf1},0)\oplus \gray{({\bf1},{\bf1},-1)} \oplus ({\bf1},{\bf1},-2) \oplus ({\bf 3},{\bf1},-1/3) \oplus ({\bf 3},{\bf1},-4/3) \oplus \gray{({\bf 3},{\bf1},+2/3)} \nonumber \\
 && \oplus \; ({\bf 6},{\bf1},+1/3) \oplus ({\bf6},{\bf1},-2/3) \oplus ({\bf 6},{\bf1},+4/3) , 
\nonumber \\
({\bf 15},{\bf 1},{\bf 1}) &=&  ({\bf1},{\bf1},0)\oplus ({\bf 8},{\bf1}, 0) \oplus \gray{({\bf 3},{\bf1},+2/3)}\oplus \gray{({\bf \overline{3}},{\bf1},-2/3)}, 
\nonumber \\
({\bf 1},{\bf 1},{\bf 3}) &=&  ({\bf1},{\bf1},0) \oplus \gray{({\bf1},{\bf1}, -1)} \oplus \gray{({\bf1},{\bf1},+1)}. 
\label{eq:10decomp1}
\eea
Here, the components in gray are the would-be Nambu-Goldstone bosons absorbed by the $SU(4)_c$ and $SU(2)_R$ gauge bosons after the PS symmetry breaking and therefore excluded from the scalar mass spectrum. 
The rest of the components obtain their masses proportional to the VEVs of their representative field.  
For simplicity, we assume that all the massive scalars contained in $({\bf 10},{\bf 1},{\bf 3})$ have approximately degenerate masses denoted by  $m_{126}$. 
We also assume that the scalars contained in the $({\bf 15},{\bf 1},{\bf 1})$ and $({\bf 1},{\bf 1},{\bf 3})$ components of $45_H$ have approximately degenerate masses denoted by $m_{45}$.

The decomposition of $10_{E}$ fermion in Eq.~(\ref{eq:PSdecomp}) under the SM gauge group is given by 
\bea
{10}^{(i=1,2)}_{E} = D^{(i)} \; ({\bf1},{\bf2},+1/2) \oplus {\bar D}^{(i)} \; ({\bf1}, \bar{\bf2}, -1/2) \oplus T^{(i)} \; ({\bf3},{\bf1}, -1/3) \oplus {\bar T}^{(i)} \; ({\bar {\bf3}},{\bf1}, +1/3).  
\label{eq:10decomp}
\eea 
We next discuss the mechanism to generate a mass-splitting between the doublet and triplet components of the ${\bf 10}$-plets, which is important for the viability of the DM scenario discussed in Sec.~\ref{sec:DM}.  
Following Ref.~\cite{DWms1}, we set the ${45}_H$ VEV to be $\langle { 45}_H\rangle = v_{45} \times {\rm diag} (1, 1, 1 , \epsilon , \epsilon) \times i \sigma_2$, with a non-zero real parameter $\epsilon \ll 1$.  
Therefore, its coupling with the ${\bf 10}$-plet fermions in Eq.~(\ref{eq:ExoticY}) yields the following mass matrices for the doublet and the triplet fermions: 
\bea
 {\cal L}_{\rm mass} \!\supset\!
\begin{pmatrix} 
{\bar D}^{(1)} & {\bar D}^{(2)} \end{pmatrix}  
\begin{pmatrix} 
0 & \epsilon Y_{45}^{12}v_{45} 
\\ \epsilon Y_{45}^{12}v_{45} & 0
\end{pmatrix} 
\begin{pmatrix} D^{(1)} \\ D^{(2)} \end{pmatrix} 
\!+ \!
\begin{pmatrix} {\bar T}^{(1)} & {\bar T}^{(2)} \end{pmatrix}  
\begin{pmatrix} 
0 & Y_{45}^{12}v_{45} 
\\ Y_{45}^{12}v_{45} & 0
\end{pmatrix} 
\begin{pmatrix} T^{(1)} \\ T^{(2)} \end{pmatrix}\!\!.  
\label{eq:DWmass} 
\eea

The doublet and the triplet fermion masses (defined in their mass eigenstates) are obtained by diagonalizing their respective mass matrices, namely, $m_{T}^{(1,2)} = Y_{45}^{12} v_{45}$ and $m_{D}^{(1,2)} = \epsilon \times m_{T}^{(1,2)}$, respectively. 
Here, we have chosen an appropriate phase such that their masses are real and positive. 
A large mass-splitting between the doublet and the triplet fermions can be realized for $\epsilon \ll 1$.  
%Note that $\epsilon > 0$ is necessary to avoid charged massless fermions in the mass spectrum. 
In the following, we define $m_{D}^{(1)}= m_{D}^{(2)} \equiv m_D$ and $m_{T}^{(1)}= m_{T}^{(2)} \equiv m_T$.

Let us consider the constraint on the triplet fermion masses. 
Through the GUT $SO(10)$ gauge interactions, the triplet fermions decay to the doublet fermion and SM fermion pair mediated by the off-shell GUT gauge bosons. 
The ${\bf 10}$-plet fermion also couples with ${10}_H$, which allows it to decay into the new singlet fermion (assumed to be lighter than the triplet fermion) and SM fermions pair mediated by the off-shell colored scalar in $10_H$. 
The partial decay widths for each of these processes are estimated as  
\bea
\Gamma_{GB} &\simeq& \frac{\left(4\pi \times \alpha_{GUT}\right)^2}{192 \pi^3} \frac{m_{T}^5}{M_{GUT}^4}, 
\nonumber \\
\Gamma_{HC} &\simeq& \frac{Y_{t}^2 {Y_H}^2}{192 \pi^3} \frac{m_{T}^5}{m_{HC}^4}, 
\label{eq:tautrip}  
\eea 
respectively. 
Here, $m_{HC}$ is the colored scalar mass, $\alpha_{GUT} = 1/20 - 1/40$ is the gauge coupling at the unification scale ($M_{GUT}$) as we will show later, 
$Y_{t} \simeq 1$ is the SM top Yukawa coupling, 
and the dimensionless coupling parameter ${Y_H} = \left(\sqrt{2} m_0/v_h\right)$, where $v_h = 246$ GeV is the SM Higgs VEV and $30 \lesssim m_0 [{\rm GeV}] \lesssim 55$, which we will discuss in Sec.~\ref{sec:DM}. 
Using these values, we find $\Gamma_{GB} \ll \Gamma_{HC}$ for $m_{HC} < M_{GUT}$, and the triplet fermion lifetime is estimated to be  
\bea
\tau_{T} \simeq \frac{1}{\Gamma_{HC}} \simeq \frac{192 \pi^3}{Y_{t}^2 {Y_H}^2} \frac{m_{HC}^4}{m_{T}^5} \simeq 1 \; {\rm s} \times \left(\frac{m_{HC}[{\rm GeV}]}{4.5 \times 10^{11}}\right)^4 \left(\frac{2.75 \times 10^5 }{m_{T}[{\rm GeV}]}\right)^5 \left(\frac{55}{m_0 [{\rm GeV}]}\right)^2. 
\label{eq:lifetime3E1}
\eea
If the triplet fermions decay after Big Bang Nucleosynthesis (BBN), their energetic decay products could destroy the light nuclei synthesized during BBN. 
To avoid this problem we simply require $\tau_T < 1$ second, so that 
\bea
m_T \gtrsim 2.75 \times 10^{5}\; {\rm GeV} \left(\frac{ m_{HC} [{\rm GeV}]}{2.7 \times 10^{12} \; \; {\rm GeV}}\right)^{4/5}, 
\label{eq:LBmT}
\eea
for $m_0 = 55$ GeV. 
For the numerical analysis of the gauge coupling unification in the next subsection, we set $m_{HC} =M_{GUT}$. 
We will also discuss the case with a lighter colored scalar.

%So fo $m_T \lesssim 8.2 \times 10^{-10} \; m_{HC}^{4/5}$, the triplet fermion is stable at the cosmological time scale with lifetime $\tau_T > 10^{26}$ seconds. 
%For example, $m_{HC} = M_{GUT} \sim 10^{15}$ GeV, requires $m_T \lesssim 10^3$ GeV; this bound gets stronger for a lighter colored scalar mass. 
%Such a light triplet fermion is excluded by exotic heavy isotope search, which puts a lower-bound on stable exotic colored fermion to be around $10$ TeV \cite{coloredDM}.   

%%%%%%%%%%%%%%%%%%%%%%%%%%%%%%%%%%%%%%
\subsection{Gauge Coupling Unification}
\label{sec:MassSpec}
%%%%%%%%%%%%%%%%%%%%%%%%%%%%%%%%%%%%% 

In this subsection we examine the unification of gauge couplings by solving the RG equations at the 1-loop level. 
Below the PS symmetry breaking scale, $\mu < M_{PS}$, 
 the RG equations for the SM gauge couplings are given by 
\bea
 \mu \frac{d \alpha_{1}}{d \mu} &=& \frac{1}{2\pi}\alpha_{1}^2
\left(\frac{41}{10}+ \frac{4}{5} \theta (\mu - m_{D})+ \frac{8}{15}\theta (\mu - m_{T}) + \frac{142}{15} \theta (\mu -m_{126})\right), 
 \nonumber \\
 \mu \frac{d \alpha_{2}}{d \mu} &=& \frac{1}{2\pi}\alpha_{2}^2 \left(-\frac{19}{6} + \frac{4}{3} \theta (\mu - m_{D}) \right), 
\nonumber \\
 \mu \frac{d \alpha_{3}}{d \mu} &=& \frac{1}{2\pi}\alpha_{3}^2 \left(-7+ \frac{4}{3}\theta (\mu - m_{T}) + 1\; \theta (\mu - m_{45}) + \frac{17}{6}\theta (\mu - m_{126}) \right), 
\label{eq:betafun1}  
\eea
where $\alpha_{i} = g_{i}^2/4\pi$ ($g_Y = \sqrt{3/5}\;g_1$ is the $U(1)_Y$ gauge coupling and $g_{2,3}$ are the $SU(2)_L$ and $SU(3)_c$ gauge couplings, respectively), and $\theta$ is the Heaviside function.
%,$m_{D,T}$ are the masses of doublet and triplet component in the two ${\bf 10}$-plet fermions, respectively, 
%and $m_{126, 45}$ are masses of all the relevant scalars in ${\bf 126}$ and ${\bf 45}$ representation Higgs which contribute the the SM gauge coupling running, respectively (see, Eq.~(\ref{eq:10decomp1})). 
For fixed values of the free parameters $m_{D,T,45,126}$, we solve the RG equations for SM gauge coupling with the following input values at $\mu = m_t = 172.44$ GeV \cite{Buttazzo:2013uya}: 
\bea
g_{1} (m_t) = \sqrt{5/3} \times 0.35830, \qquad
g_{2} (m_t) = 0.64779 , \qquad 
g_{3} (m_t) = 1.1666.  
\eea

After RG running, the SM gauge couplings at $\mu = M_{PS}$ are related to the gauge couplings of the PS gauge group through the following matching conditions at the tree-level, 
\bea
\alpha_{1}^{-1} (M_{PS}) = \frac{3}{5} {\alpha}_{R}^{-1} (M_{PS}) + \frac{2}{5}{\alpha}_{4}^{-1}(M_{PS}), 
\quad
 \alpha_{2} (M_{PS}) = {\alpha}_{L} (M_{PS}) ,
 \quad 
\alpha_{3} (M_{PS}) = {\alpha}_{4} (M_{PS}),  
\eea
where $\alpha_{4,L,R}$, respectively, are the $SU(4){_c}$, $SU(2)_L$, and $SU(2)_R$ gauge couplings of the PS gauge group. 
With these as initial values, we solve the following RG equations for the PS gauge coupling up to the unification scale $M_{GUT}$:  
\bea
\mu \frac{d {\alpha}_{4}}{d \mu} &=& \frac{1}{2\pi}{\alpha}_{4}^2 
\left(\frac{1}{3}\right), 
\nonumber \\
 \mu \frac{d {\alpha}_{L}}{d \mu} &=& \frac{1}{2\pi}{\alpha}_{L}^2 \left(\frac{10}{3}\right), 
\nonumber \\
 \mu \frac{d {\alpha}_{R}}{d \mu} &=& \frac{1}{2\pi}{\alpha}_{R}^2
\left(\frac{32}{3}\right). 
\label{eq:betafun2}  
\eea

%%%%%%%%%%%%%%%%%%%%%%%%%%%%%%%%
\begin{figure}[t]
\begin{center}
\includegraphics[scale =1]{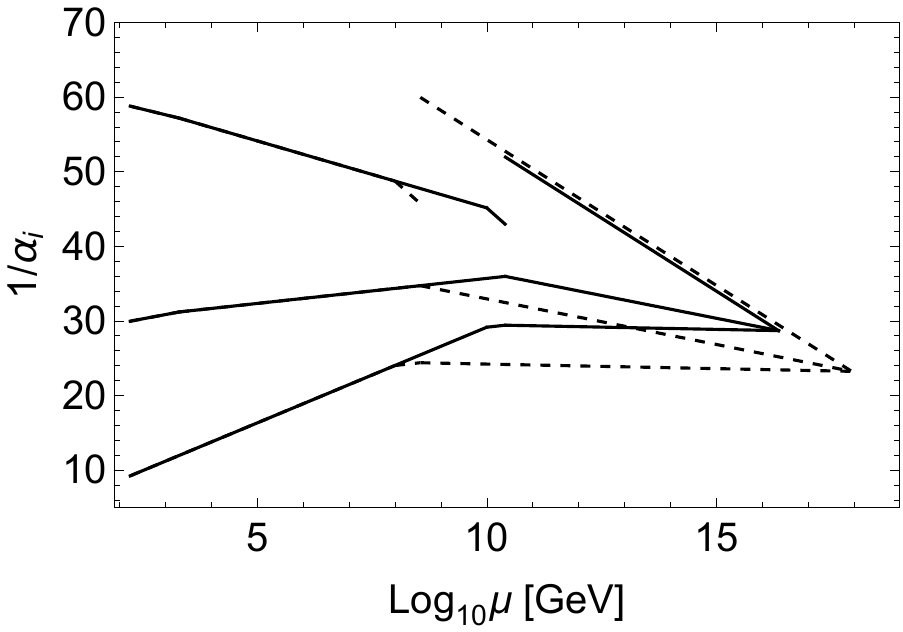} 
\end{center}
\caption{RG running of gauge couplings for $m_{D} = 2$ TeV. The solid (dashed) line denotes the result for $m_{T} = 10^5 (10^7)$ TeV. For both cases, the lines from top to bottom correspond to $\alpha_{{1,2,3}}$ for $\mu < M_{PS}$ and ${\alpha}_{{R,L,4}}$ for $M_{PS}< \mu < M_{GUT}$. 
For $m_{T} = 10^5 \; (10^7)$ TeV, we find $M_{PS} \simeq 3.5 \times 10^{8} \; ( 2.5 \times 10^{10})$ GeV and $M_{GUT} \simeq 8.8 \times 10^{17} \; (2.2 \times 10^{16})$ GeV. 
}
\label{fig:gcu}
\end{figure}
%%%%%%%%%%%%%%%%%%%%%%%%%%%%%%%%%%

The RG equations in Eqs.~(\ref{eq:betafun1}) and (\ref{eq:betafun2}) can be solved analytically as functions of  $m_{D,T,45,126}$, and $M_{PS}$.  
In addition to the gauge coupling unification conditions, namely, $ \alpha_{L} (M_{GUT}) = \alpha_{R} (M_{GUT}) = \alpha_{4} (M_{GUT}) \equiv \alpha_{GUT}$, 
we also impose the following conditions from the perturbativity of the scalar quartic couplings, $m_{D,T,45} < v_{45}\simeq M_{PQ^\prime}$, and $m_{126} < M_{PS}$. 
As an example, we plot the RG running of the gauge couplings for a fixed value of $m_{D} = 2$ TeV in Fig.~\ref{fig:gcu}, where the solid (dashed) lines correspond to $m_{T,45,126}= 10^5 \;(10^7)$ TeV.  
From top to bottom, the three solid/dashed lines denote $\alpha_{{1,2,3}}$ for $\mu < M_{PS}$ and ${\alpha}_{{R,L,4}}$ for $M_{PS}< \mu < M_{GUT}$. 
For $m_{T} = 10^5 \; (10^7)$ TeV, we find $M_{PS} \simeq 3.5 \times 10^{8} \; ( 2.5 \times 10^{10})$ GeV and $M_{GUT} \simeq 8.8 \times 10^{17} \; (2.2 \times 10^{16})$ GeV.

%%%%%%%%%%%%%%%%
\subsection{Proton Decay}
%%%%%%%%%%%%%%%

The decay $p \to \pi^0 e^+$ is mediated by the superheavy GUT gauge bosons and the colored scalars in $10_H$. 
The proton lifetime is estimated to be \cite{Nath:2006ut}
\bea
\tau_{GB} (p \to \pi^0 e^+) &\simeq& \frac{1}{\alpha_{GUT}^2} \frac{M_{GUT}^4}{m_p^5},  \nonumber \\
\tau_{HC} (p \to \pi^0 e^+) &\simeq &\frac{1}{ Y^4_d}\frac{m_{HC}^4}{m_p^5} , 
\label{eq:piondecay}
\eea  
where $m_p$ denotes the proton mass, $Y_{d} \simeq 10^{-5}$ is the down quark Yukawa coupling, and $m_{HC}$ is the colored scalar mass. 
The latter can also mediate the decay $p \to {\overline \nu}K^+$ with a lifetime 
\bea
\tau_{HC} (p \to {\overline \nu}K^+) \simeq \frac{1}{ Y^2_d Y^2_s}\frac{m_{HC}^4}{m_p^5},  
\label{eq:Kaondecay}
\eea 
 where $Y_{s} \simeq  20 \times Y_d$ is the strange quark Yukawa coupling. 
 %The non-observation of proton decay by experiments place a lower-bound on the proton lifetime. 
Applying the current Super-K bounds, $\tau_{SK} (p \to \pi^0 e^+)> 1.6 \times 10^{34}$ years  \cite{Miura:2016krn} and $\tau_{SK} (p \to {\overline \nu}K^+)> 5.9 \times 10^{33}$ years  \cite{Super-Kamiokande:2014otb}, 
we find $M_{GUT}/\sqrt{\alpha_{GUT}} > 2.5 \times 10^{16}$ GeV and $m_{HC} > 2.7 \times 10^{12} \;$ GeV, respectively. 
Combining the latter with the lifetime bound on the triplet fermion in Eq.~(\ref{eq:LBmT}), we obtain a lower-bound on the triplet fermion mass, $m_T \gtrsim 1.2 \times 10^{6} \;\rm{GeV}$.

%%%%%%%%%%%%%%%%%%%%%%%%%%%%%%%%
%\begin{figure}[!htb]
\begin{figure}[!thb]
\begin{center}
\includegraphics[scale=0.9]{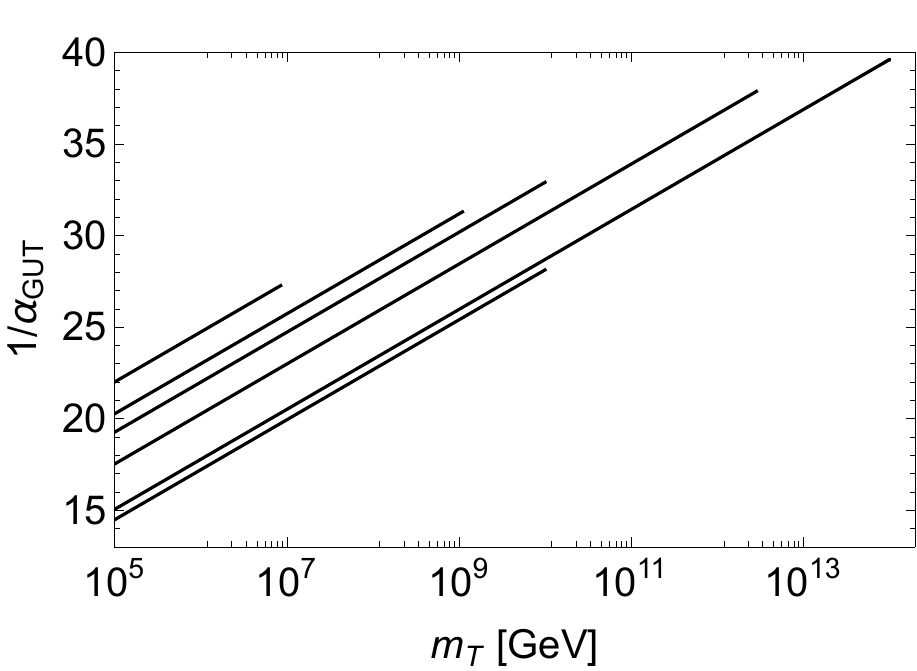}\;\;\;\includegraphics[scale=0.85]{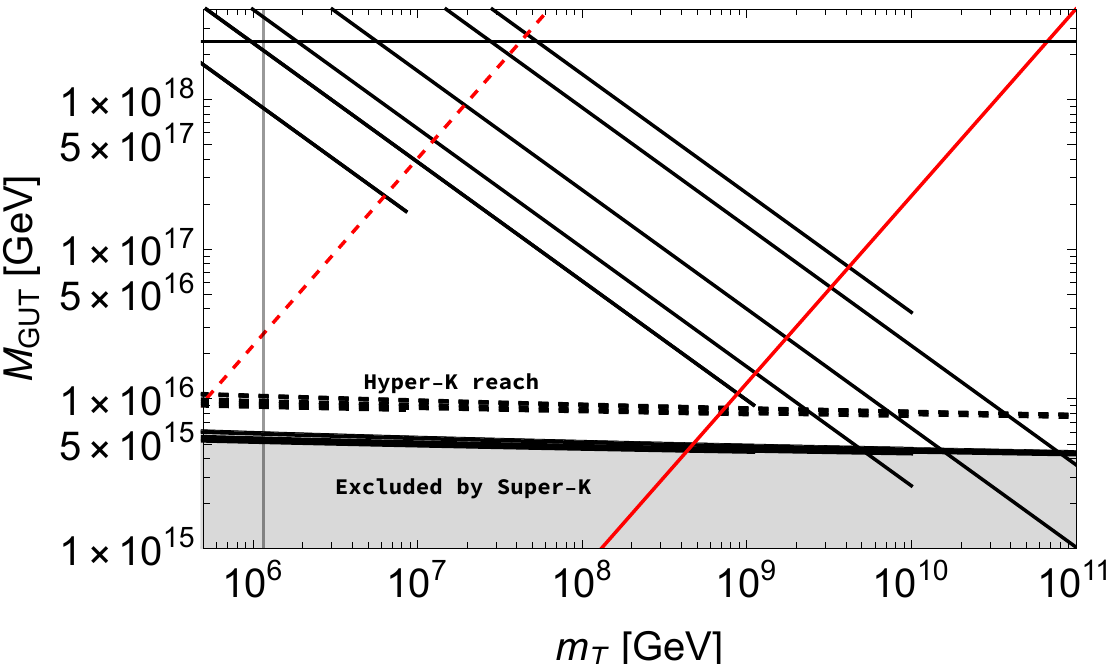}\\ \vspace{0.5cm}
\includegraphics[scale =1.0]{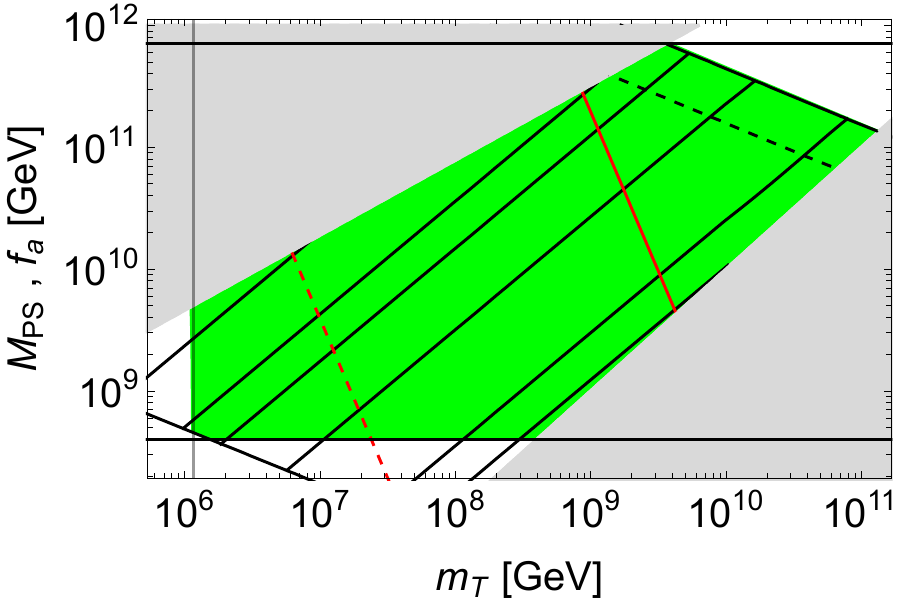}

\end{center}
\caption{Plots for $M_{PS}= f_a$ and fixed $m_D = 2$ TeV. The top-left (right) panel shows $1/\alpha_{GUT}$ ($M_{GUT}$) as a function of $m_{T}$, where the solid black diagonal lines from left-to-right (right-to-left) correspond to $m_{45,126}/m_T = 2000, 300, 100, 15, 1$, and $1/3$, respectively.  
In the top-right panel, the top horizontal line corresponds to $M_{GUT} =M_{P}$. The gray shaded region is excluded by the lower bound on proton lifetime from the Super-K experiment, and the reach of the future Hyper-K experiment is depicted by overlapping horizontal dashed lines. 
The solid black vertical line depicts the lower bound on $m_T$ for $m_{HC} = 2.7 \times 10^{12}$ GeV. 
The dashed (solid) red diagonal line depicts the lower-bound  on $m_T$ for $m_{HC} = 10^{-4} M_{GUT} \;(M_{GUT})$.
The bottom panel shows the allowed parameter region in $(m_{T}, M_{PS}=f_a$)  plane. 
The solid black diagonal lines that run from bottom-left to top-right correspond to $m_{45,126}/m_T = 2000, 300, 100, 15$, and $1$, respectively, from left to right.  
The top (bottom) solid horizontal lines depict the lower (upper) bound on $f_a = 4 \times 10^8\; (7.11 \times 10^{11})$ GeV from axion phenomenology. 
The green shaded region depicts the phenomenologically viable parameter region for $m_{HC} = 2.7 \times 10^{12} \; {\rm GeV}$. 
A heavier colored scalar mass significantly narrows down the allowed parameter space; the region to the left of the dashed (solid) red diagonal lines is excluded for $m_{HC} = 10^{-4} M_{GUT} \;(M_{GUT})$.
 }
\label{fig:MIandMGUT}
\end{figure}
%%%%%%%%%%%%%%%%%%%%%%%%%%%%%%%%%%

%%%%%%%%%%%%%%%%%%%%%%%%%%%%%%%%
%\begin{figure}[!htb]
\begin{figure}[!htb]
\begin{center}
\includegraphics[scale=0.85]{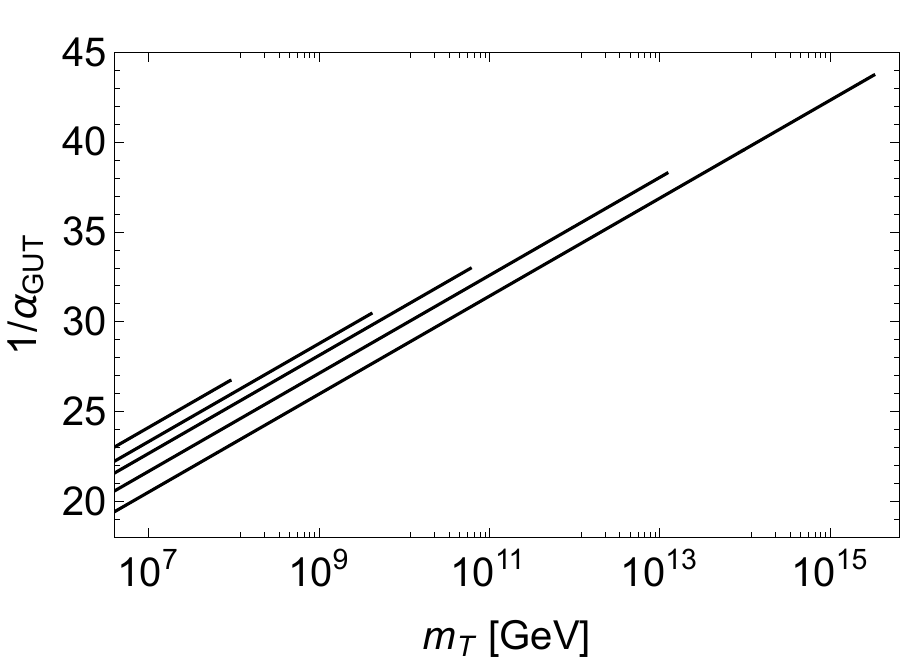}\;\;\;\includegraphics[scale=0.88]{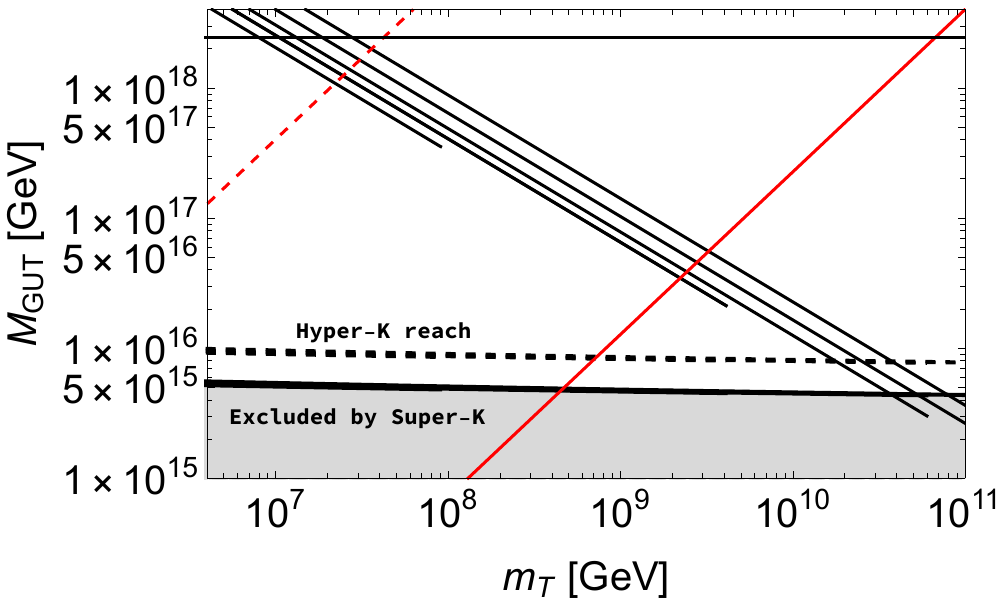}\\
\vspace{0.3cm}
\includegraphics[scale =0.9]{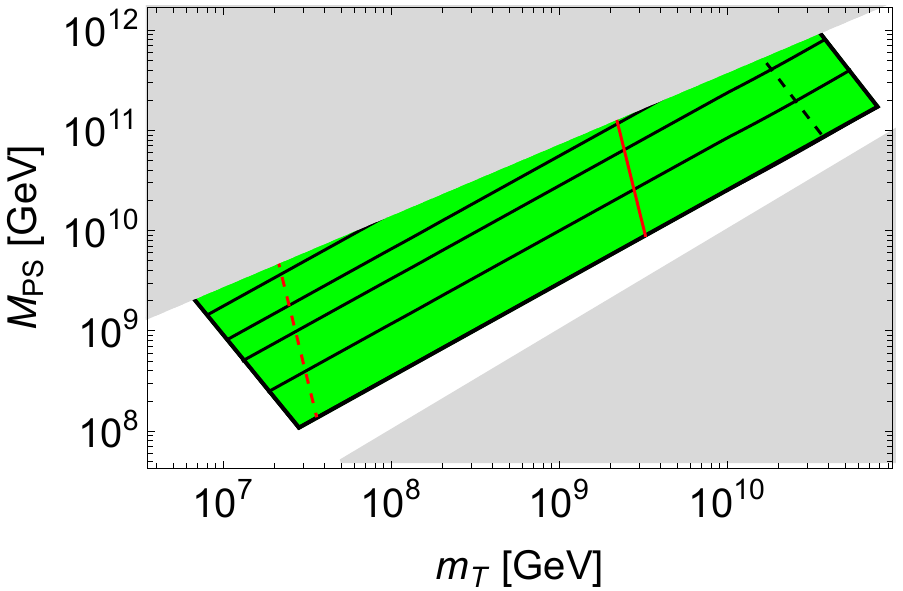}\;\;\;
\includegraphics[scale =0.9]{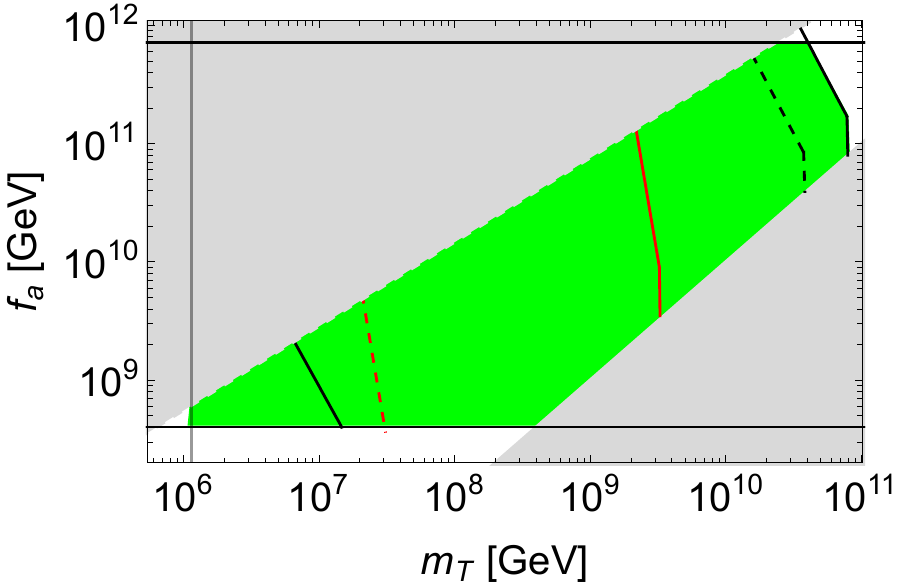}
\end{center}
\caption{
Plots for $f_a< M_{PS}$ and fixed $m_D = 2$ TeV and $m_{45} = m_T$. Top-left (top-right) panel shows $1/\alpha_{GUT}$ ($M_{GUT}$) as a function of $m_{T}$, where the solid black diagonal lines from left to right (right to left) correspond to $m_{126}/m_T = 150, 50,20$, and $1$, respectively.  
In the top-right panel, the top horizontal line corresponds to $M_{GUT} =M_{P}$. The gray shaded region is excluded by the lower bound on proton lifetime from the Super-K experiment, while the reach of the future Hyper-K experiment is depicted by overlapping horizontal dashed lines. 
The dashed and solid red diagonal lines display the lower bound on $m_T$ for $m_{HC} = 10^{-4} M_{GUT} \;(M_{GUT})$, respectively. 
The bottom left panel shows the allowed parameter region in $(m_{T}, M_{PS})$ plane. The solid black diagonal lines that run from bottom-left to top-right correspond to $m_{45,126}/m_T = 150, 50,20$, and $1$, respectively, from left to right.  
The green shaded region depicts the phenomenologically viable parameter region for $m_{HC} = 2.7 \times 10^{12} \; {\rm GeV}$. 
The solid and dashed red lines correspond to the lower bound on $m_T$ for $m_{HC} = 10^{-4} M_{GUT} \;(M_{GUT})$, respectively. The bottom-right panel shows the allowed parameter region in $(m_{T}, f_a)$ plane. 
The top (bottom) solid horizontal lines depict the lower (upper) bound on $f_a = 4 \times 10^8\; (7.11 \times 10^{11})$ GeV from axion phenomenology. 
The green shaded region depicts the phenomenologically viable parameter region for $m_{HC} = 2.7 \times 10^{12} \; {\rm GeV}$.
The solid and dashed red and black lines correspond to the lower bound on $m_T$ for $m_{HC} = 10^{-4} M_{GUT} \;(M_{GUT})$, respectively. }
\label{fig:MIandMGUT1}
\end{figure}
%%%%%%%%%%%%%%%%%%%%%%%%%%%%%%%%%%

%%%%%%%%%%%%%%%%
\subsection{Results}
%%%%%%%%%%%%%%%

%In the following discussion, we identify the phenomenologically viable parameter regions of the model. 
The gauge symmetry breaking pattern we have considered allows for two possible VEV hierarchies, namely, 
$f_a = M_{PS}$ and $f_a < M_{PS}$, which we 
separately consider. 
In Fig.~\ref{fig:MIandMGUT} we show the results for $f_a = M_{PS}$ using the benchmark values $m_D = 2$ TeV and $m_{45} = m_{126}$ (for simplicity). 
The top-left (right) panel of Fig.~\ref{fig:MIandMGUT} shows a plot of $1/\alpha_{GUT}$ ($M_{GUT}$) versus $m_{T}$, where the black solid diagonal lines from left-to-right (right-to-left) correspond to $m_{45,126}/m_T = 2000, 300, 100, 15, 1$, and $1/3$, respectively.  
For each value of $m_{45,126}/m_T$, the plot shows that there is an upper bound on $m_T$ for successful gauge coupling unification.  
Requiring $m_{D,T,45} < f_a$ and $m_{126} < M_{PS}$ only allows the mass-ratio in the range of $0.2 \lesssim m_{45,126}/m_T \lesssim 10^4$. 
For proton decay mediated by the GUT gauge bosons, the gray shaded region in the top-right panel is excluded by the Super-K result, while the reach of the future Hyper-Kamiokande (Hyper-K) experiment, 
 $\tau_{HK} \simeq 10 \times \tau_{SK}$ is depicted by horizontal dashed lines \cite{Abe:2011ts}. 
The lines depicting proton decay bounds corresponding to different $m_{45,126}/m_T$ values are almost degenerate.  
The parameter region above the top horizontal black line, $M_{GUT} > M_{P}$, is excluded. 
Although we have set $m_{HC} =M_{GUT}$ in the above analysis, we have checked that our results for gauge coupling unification very well approximate the results also for $2.7 \times 10^{12} \; {\rm GeV} < m_{HC} < M_{GUT}$ because the colored scalar contribution to the gauge coupling beta-functions is small compared to the sum of other particle contributions. 
However, lowering the colored scalar mass does change the lower-bound on $m_T$ in Eq.~(\ref{eq:LBmT}). 
For example, the solid black vertical line, the dashed red diagonal line and the solid red diagonal line depict the lower-bound on $m_T$ for $m_{HC} =2.7 \times 10^{12} \;$ GeV, $10^{-4} M_{GUT}$, and $M_{GUT}$, respectively. 
Therefore, a heavier colored scalar mass significantly narrows down the allowed parameter space.

The bottom panel shows the allowed parameter region in the $(m_{T}, M_{PS}/f_a$) plane. The solid black diagonal lines that run from bottom-left to top-right correspond to $m_{45,126}/m_T = 2000, 300$, $100, 15$, and $1$, respectively, from left to right.  
The top (bottom) solid horizontal lines depict the lower (upper) bound on the $f_a = 4 \times 10^8\; (7.11 \times 10^{11})$ GeV from axion phenomenology. 
The green shaded region depicts the phenomenologically viable parameter region for $m_{HC} = 2.7 \times 10^{12} \; {\rm GeV}$. 
A heavier colored scalar mass significantly narrows down the allowed parameter space. 
The dashed (solid) red diagonal line depicts the lower-bound on $m_T$ for $m_{HC} = 10^{-4} M_{GUT} \;(M_{GUT})$.

We show the results for $f_a < M_{PS}$ in Fig.~\ref{fig:MIandMGUT1} for benchmark values, $m_D = 2$ TeV and $m_{45} = m_{T}$ (for simplicity). 
The top-left (top-right) panel shows $1/\alpha_{GUT}$ ($M_{GUT}$) as a function of $m_{T}$, where the solid black diagonal lines from left to right (right to left) correspond to $m_{126}/m_T = 150, 50,20$, and $1$, respectively.  
In the top-right panel, we exclude the region $M_{GUT} > M_{P}$. The gray shaded region in the top-right panel is excluded by the lower-bound on proton lifetime from the Super-K experiment. 
The dashed and solid red diagonal lines are the lower bounds on $m_T$ for $m_{HC} = 10^{-4} \; M_{GUT} \;(M_{GUT})$, respectively. 
A heavier colored scalar mass significantly narrows down the allowed parameter space. 
The bottom left panel shows the allowed parameter region in $(m_{T}, M_{PS})$ plane. The solid black diagonal lines that run from bottom-left to top-right correspond to $m_{45,126}/m_T = 150, 50,20$, and $1$, respectively, from left to right.  
The green shaded region depicts the phenomenologically viable parameter region for $m_{HC} = 2.7 \times 10^{12} \; {\rm GeV}$. 
The solid and dashed red lines correspond to the lines in the top-right panel. 
The bottom right panel shows the allowed parameter region in the $(m_{T}, f_a)$ plane. 
The top (bottom) solid horizontal lines depict the lower (upper) bound $f_a = 4 \times 10^8\; (7.11 \times 10^{11})$ GeV from axion phenomenology. 
The solid and dashed red lines correspond to the same lines in the top-right panel. The green shaded region depicts the phenomenologically viable parameter region for $m_{HC} = 2.7 \times 10^{12} \; {\rm GeV}$.

%%%%%%%%%%%%%%%%%%%%%%%%%%
%%%%%%%%%%%%%%%%%
\section{Dark Matter Candidates}
\label{sec:DM}
%%%%%%%%%%%%%%%%%
%%%%%%%%%%%%%%%%%%%%%%%%%%
There are two DM candidates in our model, namely, the axion and the singlet-doublet mixed state (fermion). 
The total DM abundance is required to be the measured relic DM abundance by Planck, $\Omega_{DM} h^2 = 0.120 \pm 0.0012$ \cite{Aghanim:2018eyx}. 
In the following, we will separately evaluate each of these contributions.

%%%%%%%%%%%%%%%%%%%%%%%%%%%%%%%%%%%%%%
\subsection{Axion Dark Matter}
\label{sec:ADM}
%%%%%%%%%%%%%%%%%%%%%%%%%%%%%%%%%%%%%%
After the QCD phase transition, the axion field rolls down to the potential minima around which it oscillates coherently \cite{Kawasaki:2013ae}. 
This oscillating mode behaves like cold dark matter and contributes to the observed DM abundance such as  \cite{Kawasaki:2013ae}   
\bea
\Omega_a h^2 \simeq 0.12 \; \theta_m^2 \left(\frac{f_a}{7.11 \times 10^{11}\; {\rm GeV}}\right)^{1.19}, 
\label{eq:DMab}
\eea
where the natural choice for the misalignment angle $\theta_m \simeq 1$, and $f_a \simeq v_{45}$ is the axion decay constant. 
The results in Fig.~\ref{fig:MIandMGUT} ($f_a = M_{PS}$) and Fig.~\ref{fig:MIandMGUT1} ($f_a < M_{PS}$) show that the allowed parameter space prefers $f_a < 7.11 \times 10^{11}$ GeV. 
Therefore axion DM most likely only accounts for a fraction of the total DM abundance of $\Omega_{DM} h^2 = 0.12$.  
As an example, for the maximum $f_a$ value accessible at the future Super-K experiment, namely, 
$f_a = 2.93 \times 10^{11}$ GeV (for $f_a= M_{PS}$)
and 
$f_a= 5.39 \times 10^{11}$ GeV (for $f_a< M_{PS}$), Eq.~(\ref{eq:DMab}) indicates the axions can only account for around $10$\% and $30$\% of the observed DM abundance, respectively. 
We will now show that the remaining abundance of DM can be accounted for by the singlet-doublet fermion DM.

%%%%%%%%%%%%%%%%%%%%%%%%%%%%%%%%%%%%%%
\subsection{Singet-Doublet Fermion Dark Matter}
\label{sec:SDDM}
%%%%%%%%%%%%%%%%%%%%%%%%%%%%%%%%%%%%%%
From Eqs.~(\ref{eq:ExoticY}) and (\ref{eq:10decomp}), the Yukawa terms involving the $SU(2)_L$ doublets in the ${\bf 10}$-plets and the singlet fermion are given by 
\bea
 {\cal L} \supset \sum_{i\neq j} m_{D}^{(ij)} D^{(i)} {\bar D}^{(j)}
+ m_S 1_E 1_E + 
\sum_{i} {Y_H}^{(i)} \cos\beta \; 1_E D^{i} {H}^\dagger 
+ \sum_{j} {Y_H}^{(j)} \sin\beta \; 1_E {\bar D}^{(j)} H, 
\eea
where $m_{D}^{(12)} = \epsilon Y_{45}^{12} v_{45}$ and $m_S \equiv Y_{\Phi} v_{\Phi}$ are the doublet and singlet mass terms, respectively.  
The last two terms which mix the singlet and doublet states arise from their Yukawa interactions with $10_H \supset ({\bf 1},{\bf 2},{\bf 2}) = H_u \; ({\bf 1},{\bf 2}, +1/2) \oplus H_d \; ({\bf 1},{\bf 2}^*, -1/2)$,  
where $H_{u,d}$ contain the SM Higgs doublet ($H$), $H_{u} \supset H \sin\beta$ and $H_{d} \supset { H}^\dagger \cos\beta$, and $\tan\beta = v_u/v_d$ is the ratio of $H_u$ and $H_d$ VEVs. 
Using the choice $m_D^{(1)} = m_D^{(2)} = m_D$ from  Sec.~\ref{sec:GCUandPD} and setting $Y_H^{1} = Y_H^{2}$ (for simplicity), we obtain 
\bea
 {\cal L} \supset m_{D} D^{(1)} {\bar D}^{(2)} + m_S 1_E 1_E + {Y_H} \left( \cos\beta D^{(1)} {H}^\dagger + \sin\beta {\bar D}^{(2)} H \right) 1_E + {\rm h.c}\; .
\label{eq:YukawaDM}
\eea
Defining $D^{(1)} \equiv D$, $D^{(2)} \equiv {\bar D}$ and $1_E \equiv S$, this Lagrangian is the same as the one obtained in Ref.~\cite{IPISO10}. 
For the remainder of this section, we use their notation setting $H = 1/\sqrt{2} (0, h + v_h)^T$, where $h$ is the SM Higgs boson and $v_h = 246$ GeV. 
The mass matrix for the electrically neutral fermions, namely, $D_0$, ${\bar D_0}$, and $S$, takes the form
\bea
%{\cal L} \supset 2 m_{T} D_1 D_2 + m_S S S + m_1 D2 S + m_2 D_1 S 
\begin{pmatrix} D_0 & {\bar D_0} & S \end{pmatrix}  
\begin{pmatrix} 
0 & m_{D} & m_0 \sin\beta
\\ m_{D} & 0 & m_0 \cos\beta
\\ m_0 \sin\beta & m_0 \cos\beta & m_S
\end{pmatrix} 
\begin{pmatrix} 
 D_0 \\ {\bar D_0} \\ S
\end{pmatrix}, 
\label{eq:massmat} 
\eea 
where $m_0 \equiv {Y_H} v_h/\sqrt{2}$.

%%%%%%%%%%%%%%%%%%%%%%%%%%%%%%%
\begin{figure}[!t]
\begin{center}
\includegraphics[scale =1]{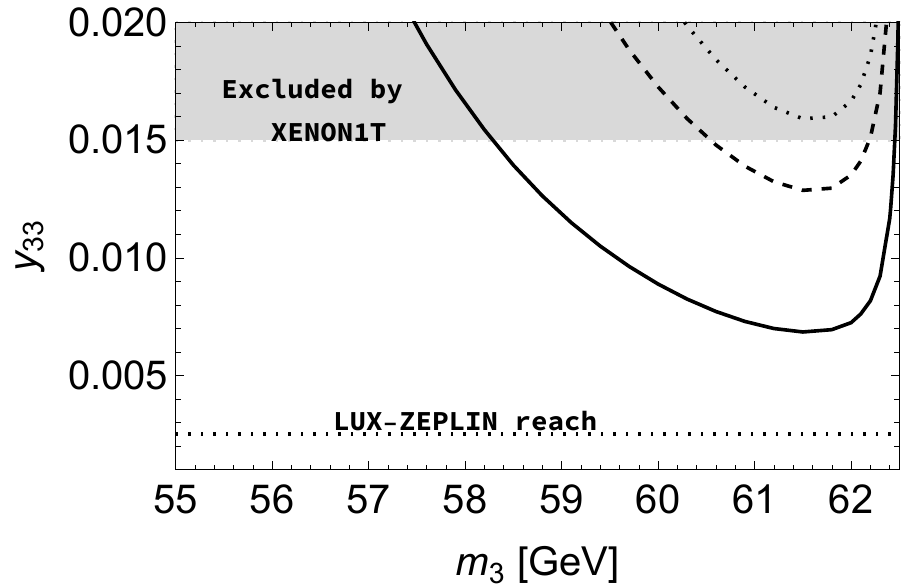}\\ \vspace{1 cm}
\includegraphics[scale=0.85]{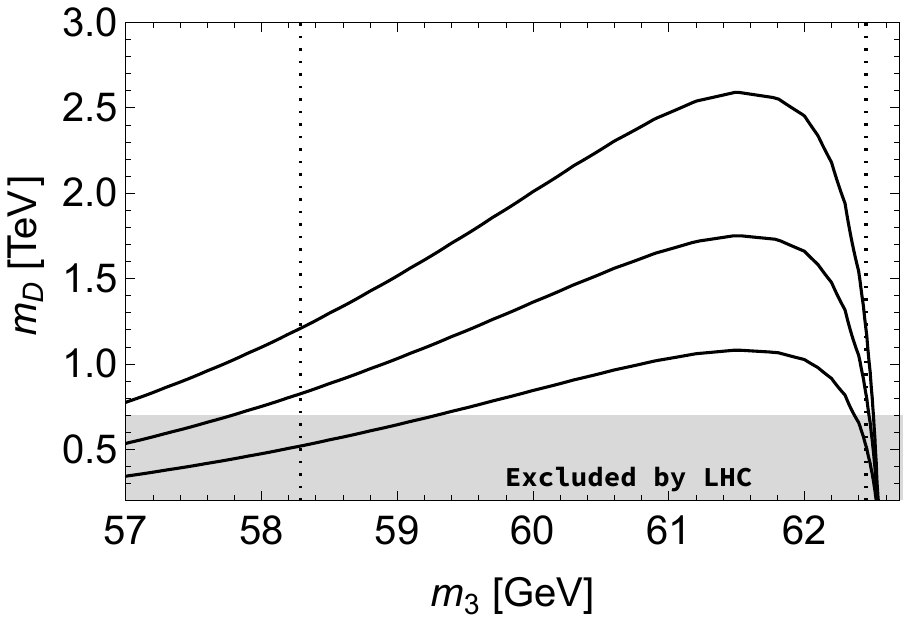} \;
\includegraphics[scale=0.85]{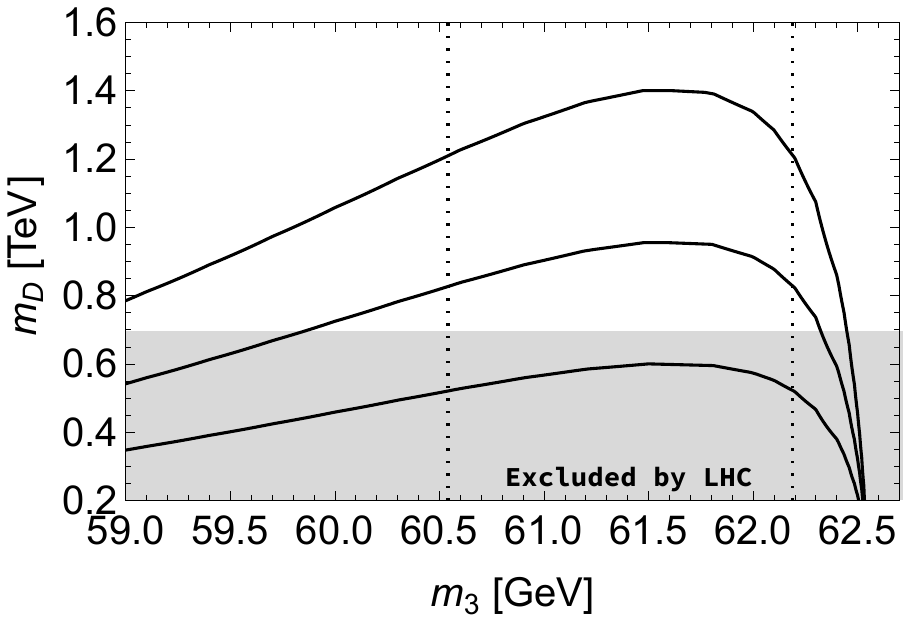}    
\end{center}
\label{fig:DM}
\caption{
For DM to be mostly the  $SO(10)$ singlet fermion, in the top-left panel we plot the effective DM Higgs portal coupling $y_{33}$ as a function of its mass $m_{3}$. 
The solid, dashed, and dotted lines are contours along which the DM accounts for $100\%, 30\%$, and $20\%$ of the observed DM abundance $\Omega_{DM} h^2 = 0.12$ \cite{Aghanim:2018eyx}, respectively. 
The gray shaded region is excluded by the XENON1T results and the search reach of the future LUX-ZEPLIN experiment is depicted by the horizontal dashed line. 
Both $y_{33}$ and $m_3$ are determined as a function of free parameters $m_0$, $\beta$ and $m_D$. 
In the bottom-left panel, for the case where the DM accounts for $100\%$ of the observed DM, we plot $m_D$ as a function of $m_3$ for fixed $\beta= \pi/3$ and for different choices of $m_0 [{\rm GeV}] = 55, 45$ and $35$ (solid curves from top to bottom)
The gray shaded region is excluded by the null LHC search results for a heavy charged lepton. 
The results for the case where DM accounts for only $30\%$ of the observed DM are shown in the bottom-right panel. Its line codings are the same as those of the bottom-left panel.  
}
\label{fig:DM}
\end{figure}
%%%%%%%%%%%%%%%%%%%%%%%%%%%%%%%%%%

We consider two extreme cases for the singlet-doublet DM scenario, ${m_S \ll m_{D}}$ and ${ m_S \gg m_{D}}$, where the DM is mostly the singlet and the doublet component, respectively. 
For ${m_S \ll m_{D}}$, the DM interacts with the SM particles effectively through its coupling with the SM Higgs boson. This scenario is well-known as the Higgs-portal fermion DM scenario. 
In the mass-basis ($\psi_{1}, \psi_{2}, \psi_{3}$), the Higgs-portal interaction is given by \cite{IPISO10}
\bea
{\cal L} &\supset& \frac{1}{2}
\begin{pmatrix} \psi_1 & \psi_2 & \psi_3 \end{pmatrix} 
U^T
\begin{pmatrix} 
0 & 0 & \frac{m_0 \sin\beta}{v_h} h
\\ 0 & 0 &  \frac{m_0 \cos\beta}{v_h}  h
\\ \frac{m_0 \sin\beta}{v_h} h & \frac{m_0 \cos\beta}{v_h}  h & 0
\end{pmatrix} 
U
\begin{pmatrix} 
 \psi_1 \\ \psi_2 \\ \psi_3
\end{pmatrix}, 
\label{eq:Lint} 
\eea 
where $U$ is the unitary matrix which diagonalizes the mass matrix in Eq.~(\ref{eq:massmat}), and $\psi_{1,2,3}$ are the mass eigenstates with masses \cite{IPISO10}
\bea
m_{1,2} \simeq m_D, \qquad m_{3} \simeq m_S - m_0 \left(\frac{m_0}{m_{D}}\right) \sin2\beta. 
\label{eq:DMmass}
\eea
The DM particle is identified with $\psi_3$, whose interactions are expressed as  
\bea 
{\cal L} &\supset & \frac {1}{2} y_{33} h \psi_{3} \psi_{3} + \frac{1}{2} y_{31} h \psi_{3} \psi_{1} + \frac{1}{2} y_{32} h \psi_{3} \psi_{2}. 
\label{eq:Lint} 
\eea 
Since the mass eigenstates $\psi_{1,2}$ are much heavier than the DM particle, the DM abundance is essentially determined by its coupling with the SM Higgs boson $h$. 
It is well-known that the allowed parameter space for the Higgs-portal DM scenario lies close to the Higgs resonance value, $m_3 \simeq m_h/2$. 
We show our results for the relic abundance calculation in the top panel of Fig.~\ref{fig:DM}. 
The solid, dashed, and dotted lines are contours along which the DM respectively accounts for $100\%, 30\%$, and $20\%$ of the observed DM abundance $\Omega_{DM} h^2 = 0.12$. 
The top gray shaded region is excluded from XENON1T \cite{XENON:2018voc} which shows that a mostly singlet DM scenario is ruled out if its relic abundance is less than $\sim 20\%$ of the observed DM abundance. 
For the solid (dashed) line, 
the allowed range of DM mass is $58.3 \lesssim m_{3}[{\rm GeV}] \lesssim 62.5$ ($60.5 \lesssim m_{3}[{\rm GeV}] \lesssim 62.2$). 
The next generation experiment LUX-ZEPLIN (LZ) will be able to probe the spin-independent (SI) cross-section, $\sigma_{\rm SI} \leq 2.8\times 10^{-12}$ pb \cite{Akerib:2018lyp}, which is an order of magnitude improvement over the current bound from XENON1T. 
The SI cross section in our case is given by $\sigma_{\rm SI} \simeq 4.47 \times 10^{-7} {\rm pb} \times y_{33}^2$ \cite{IPISO10}, which means that the LZ experiment is sensitive to a DM coupling as small as $y_{33} \simeq 2.51 \times 10^{-3}$ (the horizontal dashed line in Fig.~\ref{fig:DM}). 
Therefore, all of the currently allowed parameter regions of this DM scenario will be tested by the LZ experiment.

From Eqs.~(\ref{eq:DMmass}) and (\ref{eq:Lint}), the quantities $m_3$ and $y_{33}$ are determined in terms of the free parameters $m_0$, $\beta$ and $m_D$. 
In the bottom-left panel of Fig.~\ref{fig:DM}, we show $m_D$ as a function of $m_3$ for various $m_0$ values.  
From top-to-bottom the solid curves correspond to $m_0 [{\rm GeV}] = 55, 45$ and $35$. 
Along these curves, the singlet DM accounts for $100\%$ of the observed DM corresponding to the solid curve in the top panel of Fig.~\ref{fig:DM}. 
Here, we have considered $m_0 < 60$ GeV, for which 
the doublet fermion contribution to the beta-function of the SM Higgs quartic coupling can stabilize the SM Higgs potential as shown in Ref.~\cite{IPISO10}. 
The allowed mass range which is consistent with the XENON1T bound discussed above is depicted by the vertical dotted lines.  
The gray shaded region $m_D < 690$ GeV 
 is excluded by the CMS search results for a heavy charged lepton at the LHC \cite{CMS:2018cgi}. 
In the bottom-right panel of Fig.~\ref{fig:DM}, 
we show the results for the case where the singlet DM accounts for $30\%$ of the observed DM (dashed line in the top panel of Fig.~\ref{fig:DM}). 
The line codings for the figure are the same as the bottom-left panel of Fig.~\ref{fig:DM}. 
We find that $m_D$ must be within a few TeV for $m_0 < 60$ GeV.

Finally, the case that the DM is mostly composed of the doublet component, or equivalently $m_S \gg m_D$, is essentially the same as the Higgsino-like neutralino DM scenario in the Minimal Supersymmetric Standard Model \cite{ArkaniHamed:2006mb}. 
Since the DM annihilation cross section is determined by the electroweak gauge interactions, the DM relic abundance is only determined by the DM mass and the observed DM abundance is reproduced with $m_D \simeq 1$ TeV \cite{ArkaniHamed:2006mb}.

%%%%%%%%%%%%%%%%%%%%%%%%%%%%%%%%%%%
\section{Inflation, Reheating and Resonant Leptogenesis}
\label{sec:inf}
%%%%%%%%%%%%%%%%%%%%%%%%%%%%%%%%%%%

To solve the monopole problem, we have identified the SM singlet component of $4-2-2$ breaking Higgs in the ${126}$ representation to be the inflaton field ($\Phi$), namely, $\Phi \equiv  (1,1,0) \subset (10,1,3)$ scalar in Eq.~(\ref{eq:10decomp1}). 
For the validity of the slow-roll inflation scenario driven by a single scalar field, we will assume that the mixed couplings between the ${\overline {126}}_H$ and other Higgs fields are sufficiently small. 
We also introduce a non-minimal gravitational coupling of the $\Phi$ field with the scalar curvature.

Let us first summarize the setup for our inflation scenario.  
The action for the inflation field $\phi=\sqrt{2} {\rm Re}[\Phi]$ is defined in the so-called Jordan frame. 
In Planck units (the reduced Planck mass $M_P=2.44 \times 10^{18}$ GeV is set to equal to 1), the action is expressed as 
\begin{eqnarray}
 {\cal S}_J &=& \int d^4 x \sqrt{-g} 
   \left[-\frac{1}{2} f(\phi)  {\cal R}+ \frac{1}{2} g^{\mu \nu} \left(\partial_\mu \phi \right) \left(\partial_\nu \phi \right)  - V_J (\phi) \right]  , 
\label{S_J}
\end{eqnarray}
where $f(\phi) = (1+ \xi \phi^2)$, $\xi > 0$ is a real parameter, and the inflaton potential is 
\bea
V_J (\phi)= \lambda_{126} \left(\Phi^\dagger \Phi - M_{PS}^2 \right)^2 \simeq \frac{1}{4}\lambda_{126} \phi^4,
\label{eq:infpot}
\eea
where we have neglected the inflaton VEV in the final expression because, as we will see later, the inflaton value during the inflation is much greater than $M_{PS}$. 
For $\xi = 0$, this action reduces to the case of the standard quartic potential inflaton. 
The action in the Jordan frame can be transformed into the so-called Einstein frame with the conformal transformation $f(\phi) g_{\mu \nu} =  g_{E {\mu \nu}}$, 
\begin{eqnarray}
S_E &=& \int d^4 x \sqrt{-g_E}\left[-\frac{1}{2}  {\cal R}_E +  \frac{1}{2} g_E^{\mu \nu} \left(\partial_\mu \sigma \right) \left(\partial_\nu \sigma \right)-V_E(\phi(\sigma)) \right], 
\label{S_E}   
\end{eqnarray}
where $V_E(\phi(\sigma)) = V_J (\phi)/ f(\phi)^2$, and the cannonically normalized scalar $\sigma$ (inflaton in the Einstein frame) is related to the original field $\phi$ by  
\begin{eqnarray}
\left(\frac{d\sigma}{d\phi}\right)^{2} = \frac{1+ \xi (6 \xi +1) \phi^2} {\left( 1 + \xi \phi^2 \right)^2}. 
\label{eq:sigphi}
\end{eqnarray}
Using this relation, the inflationary slow-roll parameters in the Einstein frame are listed below:  
\begin{eqnarray}
 \epsilon(\phi) &=& \frac{1}{2} \left(\frac{V_E^\prime}{V_E \; \sigma^\prime}\right)^2,   \nonumber \\
 \eta(\phi) &=& \frac{V_E^{\prime \prime}}{V_E \; (\sigma^\prime)^2}- \frac{V_E^\prime \; \sigma^{\prime \prime}}{V_E \; (\sigma^\prime)^3} ,   \nonumber \\
 \zeta (\phi) &=&  \left(\frac{V_E^\prime}{V_E \; \sigma^\prime}\right) 
 \left( \frac{V_E'''}{V_E \; (\sigma^\prime)^3}
-3 \frac{V_E'' \; \sigma''}{V_E \; (\sigma^\prime)^4} 
+ 3 \frac{V_E^\prime \; (\sigma^{\prime \prime})^2}{V_E \; (\sigma^\prime)^5} 
- \frac{V_E^\prime \; \sigma'''}{V_E \; (\sigma')^4} \right),
\label{eq:SR1} 
\end{eqnarray}
where derivatives of $V$ with respect to $\phi$ are denoted with {\it prime}.

Inflation takes place if the slow-roll parameters satisfy $\epsilon,  |\eta|, \zeta \ll 1$. 
The amplitude of the curvature perturbation ($\Delta_{\cal R}^2 $) and the number of e-folds ($N$) are expressed as 
\begin{equation} 
  \Delta_{\cal R}^2 = \left. \frac{V_E (\phi)}{24 \pi^2 \epsilon (\phi) } \right|_{\phi_I},
\qquad 
N = \frac{1}{\sqrt{2}} \int_{\phi_{\rm e}}^{\phi_I}
  d \phi  \frac{\sigma^\prime (\phi)}{\sqrt{\epsilon(\phi)}} \;,
\label{eq:SR2}
\end{equation}
where $\phi_I$ denote the inflaton field value at the horizon exit scale, and $\epsilon(\phi_E)=1$ defines the end of inflation. 
From the Planck 2018 result \cite{Planck2018}, $\Delta_\mathcal{R}^2= 2.099 \times10^{-9}$ for the pivot scale $k_0 = 0.05$ Mpc$^{-1}$ which corresponds to $\phi_I$. 
To solve the horizon and flatness problems the number of e-folds is typically set to be $N \gtrsim 50$.

\begin{figure}[!t]
\begin{center}
\includegraphics[width=0.49\textwidth, height=5.5cm]{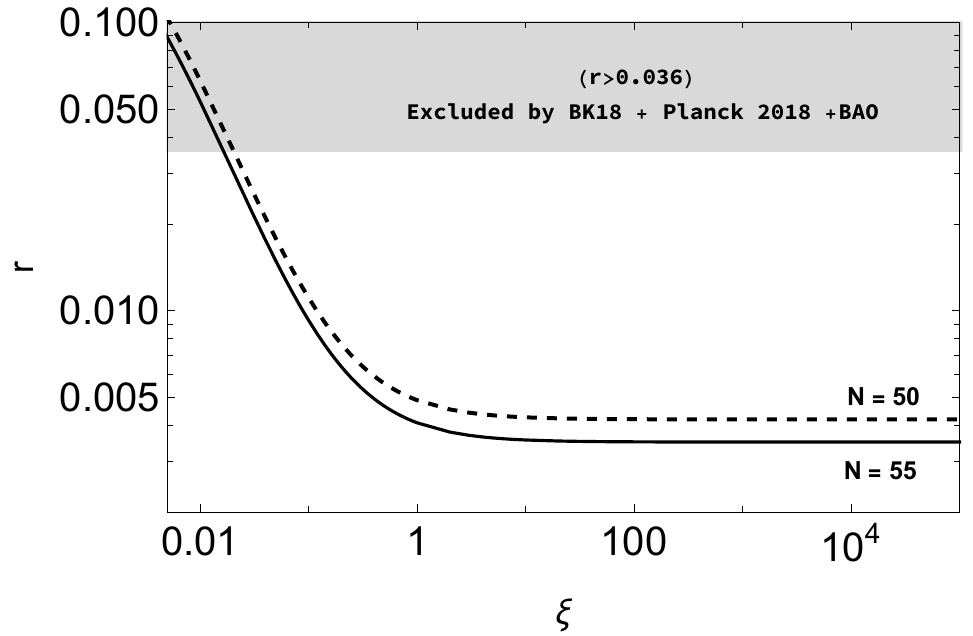} \;
\includegraphics[width=0.48\textwidth, height=5.3cm]{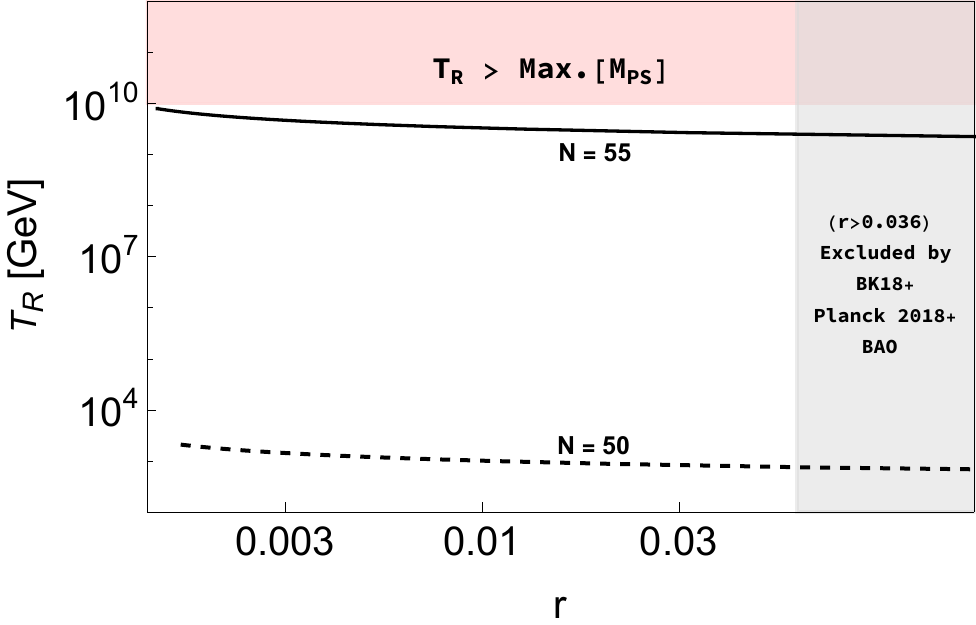} \\
\vspace{1cm}
\includegraphics[width=0.6\textwidth, height=6cm]{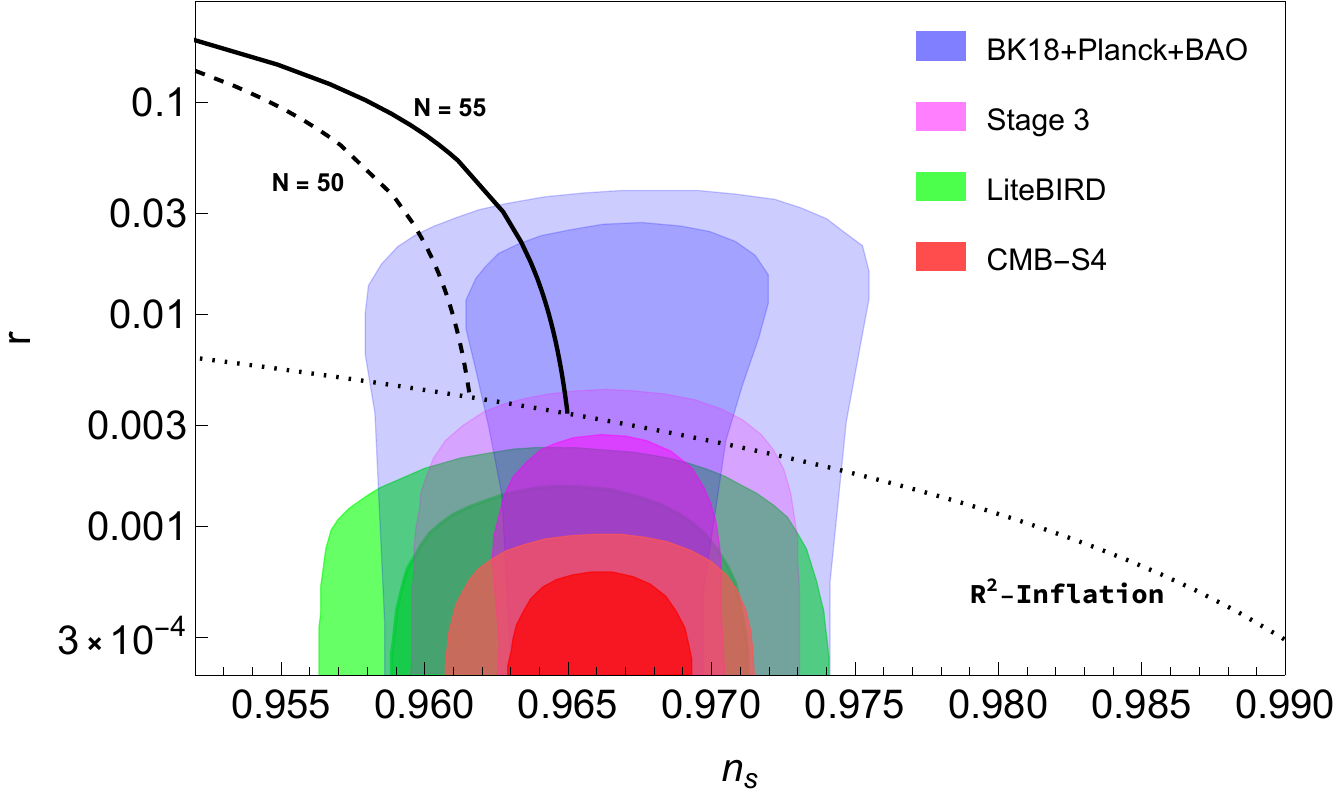} 

\end{center}
\label{fig:DM}
\caption{
The top-left panel shows $r$ as a function of $\xi$ for $N=50$ and $55$ with the shaded region excluded by the current experiments. The top-right panel shows that $N\gtrsim 55$ is excluded to prevent the restoration of PS symmetry. 
The bottom panel shows the inflationary predictions for the same N values along with the current best constraints from a combination of the BICEP2/Keck Array (BK18) experiments + Planck 2018 \cite{BICEPKeck:2022mhb} and the search reach of future experiments such as LiteBIRD \cite{LiteBIRD:2022cnt} and CMB stage 3 and stage 4 (CMB-S4) \cite{Abazajian:2019eic}. 
The dotted black line denotes inflationary prediction for $R^2$ inflation. 
}
\label{fig:infexp}
\end{figure}

The inflationary predictions for the tensor-to-scalar ratio ($r$), spectral index ($n_{s}$), and running of the spectral index ($\alpha=\frac{d n_{s}}{d \ln k}$) are given by 
\begin{eqnarray}
r = 16 \epsilon,  \;  
n_s = 1-6\epsilon+2\eta, \; 
\alpha \!=16 \epsilon \eta - 24 \epsilon^2 - 2 \zeta, 
\end{eqnarray} 
where all the three slow-roll parameters are evaluated at $\phi=\phi_I$. 
For a fixed $N$ value, substituting for $\Delta_\mathcal{R}^2$ in Eqs.~(\ref{eq:SR1}) and (\ref{eq:SR2}), the quantities
$\lambda_\Phi$, $\phi_I$ and $\phi_E$ (and therefore all the inflationary predictions) are determined as a function of the non-minimal gravitational coupling $\xi$. 
The results for $N=55$ are summarized in Table~\ref{Tab:2}.
The values of $\xi < 0.0134$ ($r > 0.036$) are excluded by the Bicep/Keck results \cite{BICEPKeck:2022mhb}. 
The inflationary predictions for $n_s, r$, and $\alpha$ rapidly approach their asymptotic values for $\xi \gtrsim 1$, for example, see Fig.~\ref{fig:infexp}. 
We also find that $\phi_I \gg M_{PS}$, which justifies the approximation used in Eq.~(\ref{eq:infpot}).

%%%%%%%%%%%%%%%%%%%%%%%%%%
\begin{table}[!t]
\begin{center}
\begin{tabular}{|c||ccc|c|}
\hline 
 $\xi $ & $n_s$   &  $r$  &  $\alpha (10^{-4})$  & $\lambda_{126} (\phi_I)$\\ 
\hline
 $0.0164$  &  $0.962$  &  $ 0.036$      & $-7.07$      &  $ 1.57\times 10^{-12}  $    \\
  $0.0745$    & $0.964$  &  $ 0.011$      & $-6.46$      &  $ 8.38  \times 10^{-12}  $    \\
  $1$             &  $0.965$  &  $0.00408$    & $-6.23$      &  $ 5.23  \times 10^{-10}  $    \\
  $10$           &  $0.965$  &  $0.00356$     & $-6.21$     &  $ 4.54  \times 10^{-8}    $    \\
  $100$         &  $0.965$  &  $0.00350$.    & $-6.21$     &  $ 4.47  \times 10^{-6}    $    \\
  $1000$       & $0.965$   &  $0.00350$    & $-6.21$      &  $ 4.46  \times 10^{-4}    $    \\
  $10^4$       &  $0.965$   &  $0.00350$   & $-6.21$      &  $ 4.46 \times 10^{-2}     $    \\
  $2\times10^4$       &  $0.965$   &  $0.00350$   & $-6.21$      &  $ 0.18$    \\
  $3\times10^4$       &  $0.965$   &  $0.00350$   & $-6.21$      &  $ 0.40$    \\
  $4\times10^4$       &  $0.965$   &  $0.00350$   & $-6.21$      &  $ 0.71 $    \\
  $4.5\times10^4$       &  $0.965$   &  $0.00350$   & $-6.21$      &  $ 0.90 $    \\
\hline
\end{tabular}
\end{center}
\caption{ 
Inflationary predictions for various benchmark $\xi$ values and $N=55$. 
Values of $\xi < 0.0134$ ($r > 0.036$) are excluded by the Bicep/Keck results \cite{BICEPKeck:2022mhb}. In our scenario, $\lambda_{126} (\phi_I) ={\cal O} (0.1)$ is required (see text for details). 
} 
\label{Tab:2}
\end{table}
%%%%%%%%%%%%%%%%%%%%%%%%%%%%%%%%%
After the end of inflation, the inflaton field rolls down to the potential minimum where it oscillates and decays to the SM particles to reheat the universe. 
For fixed values of $N$ and $\xi$, the reheat temperature $T_R$ is determined as \cite{Lyth:1998xn}\footnote{See Ref.~\cite{Kawai:2021hvs} for a more precise formula.}
\bea
N \simeq 51.4 + \frac{2}{3}{\rm ln}\left(\frac{V_E(\phi_I)}{10^{15} \; {\rm GeV}}\right) +  \frac{1}{3}{\rm ln}\left(\frac{T_R}{10^{7} \; {\rm GeV}}\right). 
\eea
If the reheat temperature $T_R > M_{PS}$, the PS symmetric vacuum gets restored during reheating and its subsequent breaking produces unacceptably large number of monopoles. 
To avoid this problem, we impose $T_R$ well below $M_{PQ}$ for the rest of our analysis, 
namely,$T_R < 10^{-2} M_{PS}$. 
In Sec.~\ref{sec:GCUandPD}, we have found the maximum value of $M_{PS} \approx 10^{12}$ GeV, 
so that the maximum reheating temperature $T_R =10^{10}$ GeV.

In Fig.~\ref{fig:infexp} we show the results for $T_R$, $n_S$ and $r$ for $N=50$ and $55$.  
The top-left panel shows $r$ as a function of $\xi$. The shaded region is excluded by the combination of the BICEP2/Keck Array (BK18) experiments and Planck 2018\cite{Abazajian:2019eic}. 
The top-right panel in Fig.~\ref{fig:infexp} shows that $N\gtrsim 55$ is excluded (gray shaded region) in order to prevent the restoration of PS symmetry.  
The bottom panel in Fig.~\ref{fig:infexp} shows the  inflationary predictions for the same $N$ values along with the current best constraints from the combination of the BICEP2/Keck Array 2018 experiments + Planck 2018 \cite{BICEPKeck:2022mhb} and the search reach of future experiments such as LiteBIRD \cite{LiteBIRD:2022cnt} and CMB stage 3 and stage 4 (CMB-S4) \cite{Abazajian:2019eic}. 
Note that the reheating bound on $N \lesssim 55$ also sets an upper bound $n_s \lesssim 0.965$.
For $\xi \gg 1$, $r$ approaches the constant value $r \simeq 3 N^2/4$ for a fixed $N$ value, and $n_s$ can be approximated as 
\bea
n_s \simeq 1-\sqrt{\frac{r}{3}}-\frac{3r}{8}. 
\eea  
The predictions for large $\xi$ is depicted by the dotted line in the bottom panel in Fig.~\ref{fig:infexp}. 
In this limit, the inflationary predictions coincide with the predictions of $R^2$ inflation \cite{Starobinsky:1980te}. 

%%%%%%%%%%%%%%%%%%%%%%%%%%%%%%%%%%
%%%%%%%%%%%%%%%%%%%%%%%%%%%%%%%%%%

The inflaton decays to the SM particles when the age of the universe equals the lifetime of the inflaton. 
This transmits the energy carried by the inflaton to the SM particles and the universe is reheated.  
Assuming instantaneous thermalization of the decay products, we estimate the reheat temperature ($T_R$) as \begin{eqnarray}
  T_R \simeq \left(\frac{90}{\pi^2 g_*}\right)^{1/4} \sqrt{\Gamma M_P},  
\label{eq:TR}
\end{eqnarray} 
where $g_*$ is the total number of thermal degrees of freedom in the plasma, and $\Gamma$ is the total decay width of the inflaton.

To evaluate the decay of the inflaton, we consider the Yukawa interaction of the inflaton with the right-handed Majorana neutrinos in the ${\bf 16}$-plet fermions. 
From Eq.~(\ref{eq:SMY}), we obtain 
\bea
{\cal L} \supset \frac{1}{4}Y_{\overline{126}}^{i} {\overline {126}}_{H} {16^i}_{SM} {16^i}_{SM} \supset \frac{1}{4}Y_{N}^{i} \Phi {\overline {N_R^{i \; C}}} N_R^{i}\;, 
\label{eq:RHNYukawa}
\eea
where we work in the flavor-diagonal basis for the right-handed neutrinos. 
%both $\Phi$ (inflaton) and $N_R^i$ (right-handed neutrino (RHN)) are the SM singlet scalar and fermion fields, respectively, and the RHN Yukawa couplings $Y_N^i$ in the last term are defined to be in the flavor-diagonal basis.  
The total decay width of the inflaton into pairs of Majorana fermions is given by
\bea
\Gamma = \Sigma_i \Gamma (\varphi \to {\overline {N_R^{i \; C}} N_R^{i} }) \simeq \frac{\Sigma_i(Y_N^{i})^2}{8\pi}m_{126}\;, 
\label{eq:gamma2}
\eea 
where $m_{126}$ is the inflaton mass. 
Both the inflaton and the Majorana fermions obtain their masses from the PS symmetry breaking. 
The Majorana mass is given by Eq.~(\ref{eq:RHNYukawa}), $m_N^i = (Y_N^i/2) M_{PS}$, and we parameterize the inflaton mass to be $m_{126} \equiv \sqrt{\lambda_{126}} M_{PS}$, where $Y_N^i$ and $\lambda_{126}$ refer to the coupling values evaluated at the PS symmetry breaking scale.

\begin{figure}[t]
\begin{center}
\includegraphics[scale=0.9]{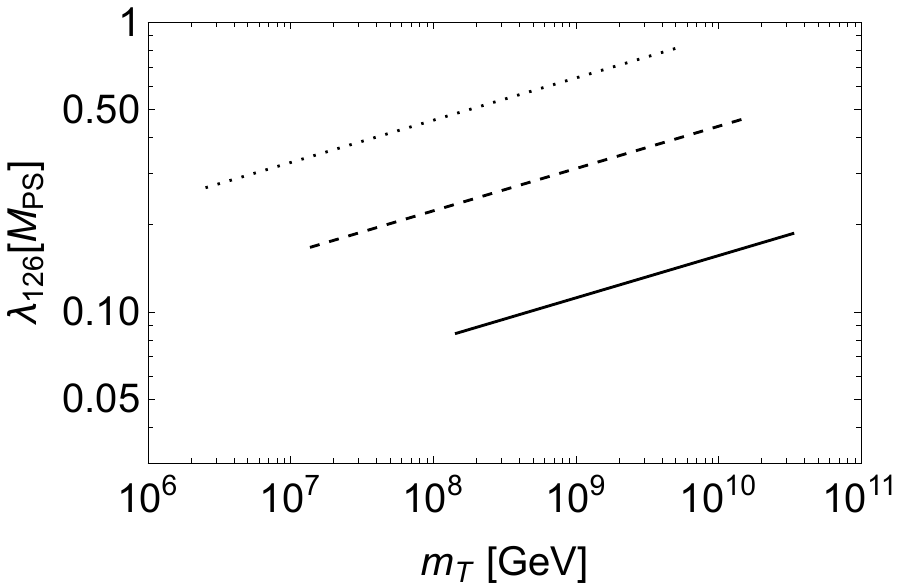}\\ \vspace{0.5 cm}
\includegraphics[scale=0.92]{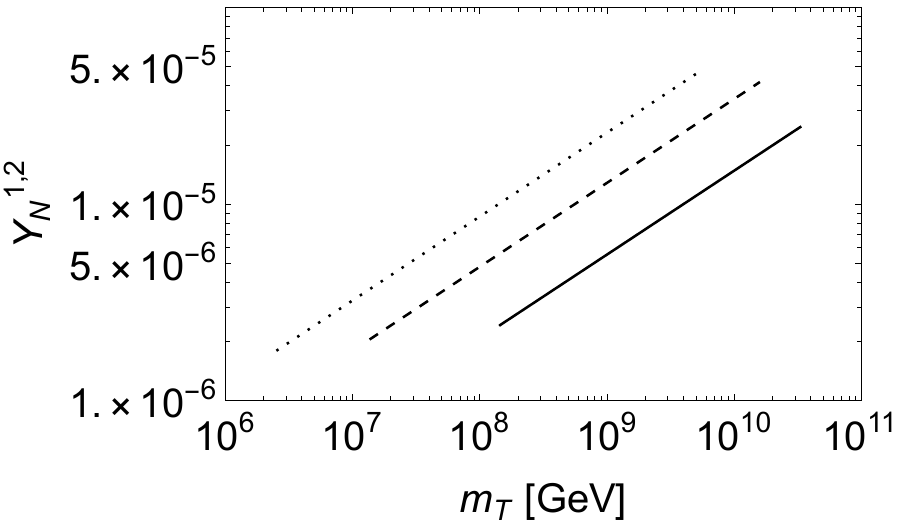}
\includegraphics[scale=0.88]{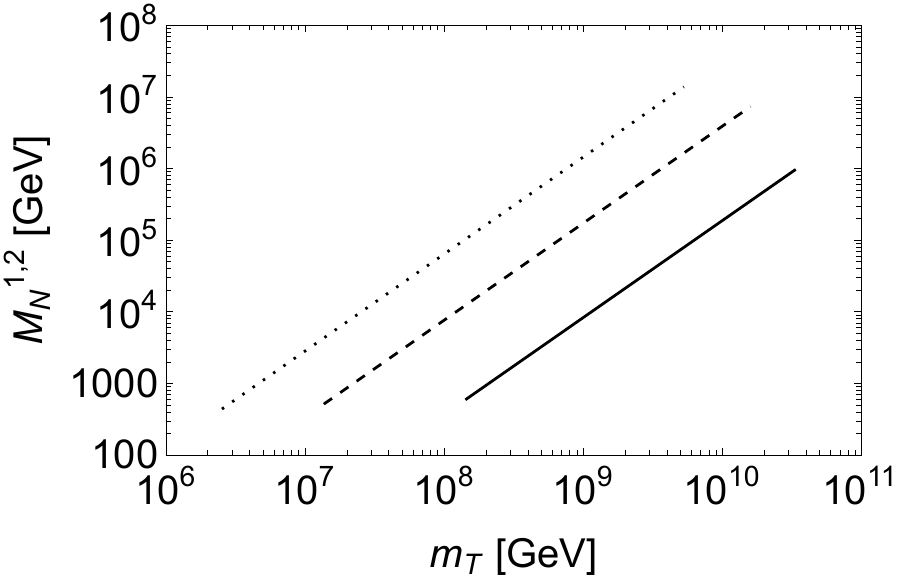}
\end{center}
\caption{For $M_{PS}= f_a$ and fixed $m_D = 2$ TeV, we show the plots for $\lambda_{126}, Y_N^{1,2}$ and $m_N^{1,2}$ as a function of $m_T$. 
The solid, dashed, and dotted lines correspond to $m_{126}/m_T = 1, 15, 100$, respectively. 
The $Y_N^{1,2}$ and $m_N^{1,2}$ correspond to their maximum values for fixed $T_R = 10^{-2} M_{PS}$, chosen to prevent the restoration of the PS symmetry. 
}
\label{fig:reheating}
\end{figure}

The inflaton mass $m_{126} = \sqrt{\lambda_{126}} M_{PS}$, together with the results for $M_{PS}$ obtained in Fig.~\ref{fig:MIandMGUT} ($f_a= M_{PS}$) and Fig.~\ref{fig:MIandMGUT1} ($f_a < M_{PS}$) for fixed values of the mass-ratio, $m_{126}/m_T$, 
allows us to determine the inflaton quartic coupling $\lambda_{126}$ as a function of $m_T$. 
Here we focus on the case $M_{PS} = v_{45}$ with $m_D^{1,2} = 2$ TeV.  
The results are shown in the top panel of Fig.~\ref{fig:reheating}, where the solid, dashed, and dotted lines correspond to $m_{126}/m_T = 1, 15, 100$, respectively. 
From the $\lambda_{126}$ values, together with Eqs.~(\ref{eq:TR}) and (\ref{eq:gamma2}), the Yukawa couplings are  determined as   
\bea
\sqrt{\Sigma_i \left(Y_N^i\right)^2} \simeq 5.8\times10^{-9} {\rm GeV}^{-1/2} \frac{T_R}{\sqrt{M_{PS}}}\frac{1}{\sqrt{\lambda_{126}}}\;. 
\label{eq:maxY}
\eea 
Using $T_R = 10^{-2} M_{PS}$, we obtain $Y_N^i \ll 1$ even for $M_{PS} \sim 10^{11}$ GeV, the maximal value allowed for a successful unification of gauge couplings. 
In the following, we assume $Y_N^{1} \simeq Y_N^{2}$ and $Y_N^3 > \sqrt {\lambda_{126}}$, such that the inflaton only decays to $N_R^{1,2}$.\! \footnote{According to fermion mass fitting analysis \cite{S010nonSUSY}, $Y_{N}^3 \simeq 1$. Note that the condition in Eq.~(\ref{eq:maxY}) is only applicable to the case $Y_N^i < \sqrt{\lambda_{126}}$.}
In the bottom left and right panels of Fig.~\ref{fig:reheating}, we show $Y_N^{1,2}$ and the corresponding Majorana fermion masses, $m_N^{1,2}$, as a function of $m_T$. 
The solid, dashed, and dotted lines in Fig. \ref{fig:reheating} correspond to $m_{126}/m_T = 1, 15, 100$, respectively. 
Note that the $Y_N^{1,2}$ and $m_N^{1,2}$ values plotted in Fig. \ref{fig:reheating} correspond to the maximum values for our benchmark $T_R = 10^{-2} M_{PS}$ to prevent the restoration of the PS symmetry.

%%%%%%%%%%%%%%%%%%%%%%%%%%%%%%%%
%\begin{figure}[!htb]
%%%%%%%%%%%%%%%%%%%%%%%%%%%%%%%%%%

Let us next discuss leptogenesis in our model. 
A successful thermal leptogenesis scenario with non-degenerate Majorana neutrinos requires the mass of the lightest right-handed neutrino to be heavier than $10^{9-10}$ GeV \cite{leptogenesis2}, with a upper bound on its mass of around $10^{15}$ GeV imposing perturbative bound on Dirac Yukawa coupling. 
Since this scenario does not work in our case with $T_R = 10^{-2} M_{PS} < 10^{9-10}$ GeV, we consider a resonant leptogenesis scenario \cite{Flanz:1996fb, Pilaftsis:1997jf} with the masses of the two lightest Majorana neutrinos nearly degenerate, which enhances the CP-asymmetry parameter.  
We therefore have set $Y_N^{1} \simeq Y_N^{2}$. To thermalize these Majorana neutrinos the reheat temperature $T_R > m_N^{1,2}$, or equivalently, $Y_N^{1,2} < 2 (T_R/M_{PS}) = 0.02$, which is consistent with the values shown in the bottom-left panel of Fig.~\ref{fig:reheating}.

In the top panel of Fig.~\ref{fig:reheating} we have shown that the inflaton quartic coupling values at the PS symmetry breaking scale, $\lambda_{126} (M_{PS}) \simeq {\cal O}(0.1)$.  
Its values during inflation, $\lambda_{126} (\phi_I)$, are listed in Tab.~\ref{Tab:2} for different choices of $\xi$. 
After taking into account the RG running of $\lambda_{126}$, we expect a one-to-one correspondence between these two values of $\lambda_{126}$.  
To establish this connection, 
we consider the RG equation for the quartic coupling of the inflaton $\Phi \equiv (1,1,0) \subset (10,1,3)$. 
At the 1-loop level, we obtain  
\bea
\mu \frac{d \lambda_{126}}{d \mu} \!\!\!\!&= &\!\!\!\! \frac{1}{16\pi^2} \left( 64 \lambda_{126}^2 + 2 \lambda_{126} \left( (Y_{N}^3)^2 -9 g_4^2 - 12 g_R^2 \right) +24 g_R^4 + \frac{27}{2} g_4^4 - \frac{1}{2} (Y_{N}^3)^4\right), 
\eea
where $g_{4,R}$ are the $SU(4)_c$ and $SU(2)_R$ gauge couplings, respectively, and we have only considered the contribution from the third generation Majorana Yukawa coupling $Y_{N}^3$ since $Y_{N}^{1,2} \ll 1$. 
To understand the RG running behavior of the quartic coupling in the range $M_{PS} \lesssim \mu \lesssim \phi_I$, note that from Figs.~\ref{fig:gcu} and \ref{fig:reheating}, $g_{4,R}^2 = 4 \pi \alpha_{4,R} \simeq \lambda_{126} \simeq {\cal O } (0.1)$ at $\mu = M_{PS}$. 
We set $Y_{N}^3 \simeq \sqrt {\lambda_{126}}$, so that the beta-function is positive and we obtain $ \mu \frac{d \lambda_{126}}{d \mu} = {\cal O } (1) \lambda_{126}^2$. 
Since $\lambda_{126} (\mu)$ remains positive for $\mu > M_{PS}$, the inflaton potential remains stable during inflation at $\mu = \phi_I$, which is crucial for the viability of the inflation scenario. 
We expect the inflaton quartic coupling during inflation to lie in the range $\lambda_{126} (\phi_I) = {\cal O} (0.1)$. 
From Table.~\ref{Tab:2}, this corresponds to $\xi = {\cal O} (10^4)$, for which the predicted $r$ can be probed by future experiments such as CMB-S4 \cite{Abazajian:2019eic}.

For completeness, a few remarks regarding iso-curvature perturbations, monopoles, and cosmic strings are in order here. The breaking of $SO(10)$ via 4-2-2 produces GUT scale and intermediate scale monopoles, respectively. The $SO(10)$ and PS symmetry are already broken during the inflation, which is driven by the PS breaking Higgs field, hence the superheavy and intermediate scale monopoles are both inflated away. 
Furthermore, since the reheat temperature after inflation is set to be $T_R  = 10^{-2} M_{PS}$, the production of the intermediate mass monopoles from thermal fluctuations is exponentially suppressed. Similarly, the intermediate scale strings are inflated away. 
For a recent discussion of a closely related $SO(10)\times U(1)_{PQ}$ model in which the strings are produced after inflation see Ref.~\cite{Lazarides:2022spe}. 
The string-wall system associated with the axion sector appears after inflation but, as we noted earlier, the domain wall problem is absent in our model by construction.

%\bea
%\lambda_{126}(\phi_I) \simeq \lambda_{126} (M_{PS}) + \frac{\beta_\lambda({\phi = M_{PS}}) }{16\pi^2} {\rm ln} \left(\frac{\phi_I}{M_{PS}}\right), 
%\eea

%%%%%%%%%%%%%%%%%
\section{Conclusions}
\label{sec:conc}
%%%%%%%%%%%%%%%%%

We have considered a simple non-supersymmetric $SO(10)$ GUT model with a global PQ symmetry $U(1)_{PQ}$. 
In addition to three fermions in the representation ${\bf 16} (+1)$ (each ${\bf 16}$-plet includes one generation of SM fermions and a SM singlet Majorana neutrino), 
we have introduced two fermions in ${\bf 10} ($-2$)$ and one in ${\bf 1} ($+4$)$ representations. 
The Majorana neutrino can explain the observed light neutrino via the type-I seesaw mechanism.

We have chosen suitable Higgs representations under $SO(10) \times U(1)_{PQ}$ to break the symmetry down to the SM via an intermediate Pati-Salam (PS) gauge group. 
Phase transition associated with the symmetry breaking can potentially produce topological defects by the Kibble mechanism, namely, domain wall from the PQ symmetry breaking and monopoles from the $SO(10)$ and PS symmetry breakings. 
Their contributions to the energy density of the universe must be suppressed to reproduce our universe. 
In the presence of the ${\bf 10}$-plets, the PQ is broken down to ${\bf Z_4}$, which coincides with the center of $SO(10)$ (more precisely, $Spin (10)$), or equivalently, $N_{DW} = 1$. 
Hence, our model is free from the axion domain-wall problem. 
The SM singlet Higgs field in the $126$ representation of $SO(10)$, whose VEV generates right-handed neutrino masses, is identified with the inflaton. 
Since the $SO(10)$ and PS symmetry are already broken during inflation, our model is also free from the monopole problem. 
Furthermore, since the PQ symmetry breaking takes place after the inflation, the isocurvature perturbations generated by the axion field are negligibly small.

Although axion is an attractive DM candidate, we find that 
it does not account for 100\% of the observed DM in the universe for most of the allowed parameter region after imposing various phenomenological constraints on the model parameter space, such as successful gauge coupling unification and proton lifetime in agreement with the current bound. 
However, the model has another compelling DM candidate, namely, a linear combination of the $SO(10)$ singlet fermion and the doublet fermions in the fermion ${\bf 10}$-plets. 
For this singlet-doublet DM scenario, we have considered two extreme cases in which the fermionic DM is either mostly the singlet fermion or the doublet fermion. 
We have identified the model parameter regions that reproduce the observed DM abundance as the total abundance from fermion and axion DMs.

We have identified the SM singlet component of the Higgs field which breaks the PS symmetry with the inflaton with non-minimal coupling to gravity. 
A successful unification of the SM gauge couplings requires the inflaton quartic coupling value to be ${\cal O} (0.1)$, which corresponds to $\xi = {\cal O} (10^4)$, the dimensionless non-minimal gravitational coupling parameter. 
For the scalar spectral index and tensor-to-scalar ratio we find $0.963 \lesssim n_s \lesssim 0.965$ and $0.003 \lesssim r \lesssim 0.036$, respectively, which can be tested by the CMB-S4 experiment.  
The inflaton decay to Majorana neutrinos reheats the universe, with the reheating temperature about two orders of magnitude below the PS symmetry breaking scale. 
With the RHN mass smaller than $10^7$ GeV, the observed baryon asymmetry in the universe is generated via resonant leptogenesis with two nearly mass degenerate right-handed neutrinos.

%%%%%%%%%%%%%%%%%%%%%%%%%%%%%%%%%%%%%%%%%
\section*{Acknowledgements}
%%%%%%%%%%%%%%%%%%%%%%%%%%%%%%%%%%%%%%%%%
Q.S would like to thank Rinku Maji for discussions. This work is supported in part by the United States Department of Energy Grants DE-SC0012447 (N.O) 
  and DE-SC0013880 (D.R and Q.S) and Bartol Research Grant BART-462114 (D.R).

%%%%%%%%%%%%%%

\end{document}